\PassOptionsToPackage{capitalize,nameinlink}{cleveref}
\documentclass[sigconf,9pt]{acmart}

\graphicspath{{images/}}
\usepackage[a-1b]{pdfx} 
\usepackage{enumitem}
\usepackage{mwe}
\usepackage{array}
\usepackage{bbm}
\usepackage{amsmath}
\usepackage{algorithmic}
\usepackage[ruled,vlined]{algorithm2e}
\usepackage{pgfplots,pgfplotstable}  
\usepackage{booktabs}
\usepackage{balance}
\usepackage{multirow}
\usepackage{subcaption}
\pgfplotsset{compat=newest}
\usepackage{balance}
\usepackage{fancybox} 
\usepackage[most]{tcolorbox}
\usepackage{lipsum}
\usepackage{graphicx}
\usepackage{cleveref}
\usepackage{appendix}
\usepackage{float}
\usepackage{xcolor}
\usepackage{xspace}
\newcommand{\embedding}{text-embedding-3-large\xspace}
\newcommand{\chat}{GPT-4o\xspace}
\newcommand{\system}{Copilot DECO\xspace}
\newcommand{\nudge}{NUDGE\xspace}
\newcommand{\framework}{\textcolor{black}{FLAIR}\xspace}
\newcommand{\companyname}{Microsoft\xspace}
\newcommand{\sqldb}{\textcolor{black}{Azure SQL DB}\xspace}
\newcommand{\sqldw}{\textcolor{black}{Microsoft Fabric}\xspace}
\newcommand{\threshold}{confidence\xspace}

\newcommand{\yz}[1]{\textcolor{brown}{[Yiwen: #1]}}

\newcommand{\sk}[1]{\textcolor{blue}{[Subru: #1]}}
\definecolor{teal}{HTML}{00AEB3}

\newcommand\redsout{\bgroup\markoverwith{\textcolor{red}{\rule[0.5ex]{2pt}{0.4pt}}}\ULon}

\definecolor{lightgray}{rgb}{.70,.70,.70}  
\definecolor{orange}{RGB}{255,127,0}

\newcommand*\circled[1]{%
    \tikz[baseline=(char.base)]{\node[shape=circle,fill=gray,draw,inner sep=0.5pt] (char) {\color{white}\textsf{#1}};}%
}%

\newtoggle{full}

\NewEnviron{outputfull}{%
  \iftoggle{full}{\BODY}{}%
}
\NewEnviron{outputabridge}{%
  \iftoggle{full}{}{\BODY}%
}

\copyrightyear{2025}
\acmYear{2025}
\setcopyright{cc}
\setcctype{by}
\acmConference[CIKM '25]{Proceedings of the 34th ACM International Conference on Information and Knowledge Management}{November 10--14, 2025}{Seoul, Republic of Korea}
\acmBooktitle{Proceedings of the 34th ACM International Conference on Information and Knowledge Management (CIKM '25), November 10--14, 2025, Seoul, Republic of Korea}\acmDOI{10.1145/3746252.3761553}
\acmISBN{979-8-4007-2040-6/2025/11}

\newcommand{\captionTitle}[1]{\textbf{#1} --}

\begin{document}
\title{\framework: Feedback Learning for Adaptive Information Retrieval}


\author{William Zhang*}
\orcid{0000-0002-9392-6683}
\affiliation{%
  \institution{Carnegie Mellon University} 
  \city{Pittsburgh}
  \country{USA}
}
\email{wz2@andrew.cmu.edu}
\thanks{*Work done while interning at Microsoft.}

\author{Yiwen Zhu}
\orcid{0009-0005-6857-7505}
\affiliation{%
  \institution{Microsoft} 
  \city{Mountain View}
  \country{USA}
}
\email{zhu.yiwen@microsoft.com}

\author{Yunlei Lu}
\orcid{0009-0005-4175-0371}
\affiliation{%
  \institution{Microsoft}
  \city{Redmond}
 \country{USA}
}
\email{yunleilu@microsoft.com}

\author{Mathieu Demarne}
\orcid{0009-0002-9841-2259}
\affiliation{%
  \institution{Microsoft}
  \city{Redmond}
  \country{USA}
}
\email{mdemarne@microsoft.com}

\author{Wenjing Wang}
\orcid{0000-0002-2011-9725}
\affiliation{%
  \institution{Microsoft}
  \city{Redmond}
  \country{USA}
}
\email{wenjing.wang@microsoft.com}

\author{Kai Deng}
\orcid{0009-0003-6832-6947}
\affiliation{%
  \institution{Microsoft}
  \city{Redmond}
  \country{USA}
}
\email{kaideng@microsoft.com}

\author{Nutan Sahoo}
\orcid{0009-0008-2545-3766}
\affiliation{%
  \institution{Microsoft} 
  \city{Cambridge}
  \country{USA}
}
\email{nutansahoo@microsoft.com}

\author{Katherine Lin}
\orcid{0009-0009-0583-7727}
\affiliation{%
  \institution{Microsoft} 
  \city{Redmond}
 \country{USA}
}
\email{katlin@microsoft.com}

\author{Miso Cilimdzic}
\orcid{0009-0004-2870-6555}
\affiliation{%
  \institution{Microsoft}
  \city{Aliso Viejo}
 \country{USA}
}
\email{misoc@microsoft.com}

\author{Subru Krishnan}
\orcid{0009-0007-8534-0889}
\affiliation{%
  \institution{Microsoft} 
  \city{Barcelona}
  \country{Spain}
}
\email{subru@microsoft.com}
\renewcommand{\shortauthors}{William Zhang et. al}

\begin{abstract}

Recent advances in Large Language Models (LLMs) have driven the adoption of copilots in complex technical scenarios, underscoring the growing need for specialized information retrieval solutions. In this paper, we introduce \framework, a lightweight, feedback learning framework that adapts copilot systems' retrieval strategies by integrating domain-specific expert feedback.
\framework operates in two stages: an offline phase obtains indicators from (1) user feedback and (2) questions synthesized from documentation, storing these indicators in a decentralized manner. An online phase then employs a two-track ranking mechanism to combine raw similarity scores with the collected indicators. This iterative setup refines retrieval performance for any query.
Extensive real-world evaluations of \framework demonstrate significant performance gains on both previously seen and unseen queries, surpassing state-of-the-art approaches. The system has been successfully integrated into \system, serving thousands of users at \companyname, demonstrating its scalability and effectiveness in operational environments.


\end{abstract}



\maketitle


\section{Introduction}

The emergence of large language models (LLMs), such as GPT~\cite{gpt}, LLaMA~\cite{llama}, and Gemini~\cite{gemini}, has revolutionized the way we interact with and manage Information Retrieval (IR) systems. By integrating Retrieval-Augmented Generation (RAG) as their backbone, these systems enable centralized and intelligent access to information, facilitating LLMs to generate accurate and contextually aware responses~\cite{ragsurvey, chen2023benchmarkinglargelanguagemodels} and supporting a diverse range of applications, often referred to as ``copilots'' (e.g., ~\cite{openaiassistant,einstein,claude}). 
Despite their transformative potential, RAG systems face significant challenges in domain-specific applications, particularly in specialized technical fields.
The primary issue stems from their reliance on embedding similarity for document ranking. These embeddings are typically trained on general-purpose corpora, such as Wikipedia~\cite{embedding}, which struggles to account for the specialized terminology or concepts prevalent in technical domains and internal information.
The resulting mismatching significantly reduces the relevance and accuracy of the retrieved information.
Additionally, for interactive copilots, as the interaction occurs in real-time, this imposes further constraints on latency, necessitating rapid and efficient retrieval mechanisms.

From observing these copilot systems' deployment at Microsoft (\cref{tab:motivation}), 
two critical observations arise. First, 
the limited
feedback from domain experts provides invaluable insights into the copilot's performance and accuracy. Second, users often submit queries that are semantically similar \textit{but not identical} to prior queries.
These insights highlight a compelling need to develop adaptive retrieval algorithms that incorporate user feedback and historical conversations to automatically enhance a copilot's abilities. 

\begin{table}[t!]
    \centering
    \resizebox{0.8\columnwidth}{!}{
    {
\begin{tabular}{lrr}
    \toprule
    & \multicolumn{1}{c}{\textbf{\sqldb}} 
    & \multicolumn{1}{c}{\textbf{\sqldw}} \\
    \midrule

    \textbf{Deployment Date} & 2024-02 & 2023-10 \\
    \textbf{Total Queries} & 1200 & 3285 \\
    \textbf{\% Queries with Feedback} & 2.4\% & 4.4\% \\
    \textbf{Unique Queries} & 643 & 1641 \\
    \textbf{$2x$ Repeating Queries} & 102 & 314 \\
    \textbf{$3x$ Repeating Queries} & 49 & 105 \\
    \bottomrule
\end{tabular}}
    }
    \caption{
        \captionTitle{Query Clusters} Clustering analysis of user queries for two copilots at \companyname using HDBScan~\cite{hdbscan}.
    }
    \label{tab:motivation}
    \vspace{-1cm}
\end{table}

Existing techniques aimed at 
generalizing across similar user queries have centered on \textbf{caching LLM outputs}~\cite{bang2023gptcache,jin2024ragcache}, \textbf{adapting embeddings}~\cite{zeighami2024nudge,zhang2024raft}, and \textbf{Learning-To-Rank}~\cite{WANG2024111334,ai2018, azizi2023overcomingpriormisspecificationonline}.
However, these methods are marred by several limitations. Caching techniques rely on brittle threshold-based similarity measures (e.g., cosine distances) to retrieve LLM outputs from a centralized cache that result in scalability challenges and suboptimal generalization. For the latter two, they face significant barriers in practical industrial applications as they require substantial supervised training data and incur additional training costs, making them impractical for frequently evolving knowledge bases and user queries. Moreover, Learning-To-Rank introduces additional models that are often not interpretable, which is a critical requirement for Copilot-style applications that must provide transparent and trustworthy results.

\subsubsection*{\textbf{Introduction to \framework}}




To address the aforementioned limitations, we propose a novel framework \textbf{\framework}---\textbf{F}eedback \textbf{L}earning for \textbf{A}daptive \textbf{I}nformation \textbf{R}etrieval. \framework leverages feedback-driven mechanisms inspired by reinforcement learning~\cite{kaelbling1996reinforcement} to iteratively refine retrieval strategies for domain-specific copilots. By doing so, \framework overcomes the brittleness and rigidity of caching solutions while also avoiding high computational overhead.
Specifically, \framework is designed with four objectives:
(1) Avoid the high computational costs associated with fine-tuning LLMs or embeddings,
(2) Adaptively generate revised answers in response to user feedback,
(3) Improve the retrieval of relevant documents for both previously seen and unseen queries, even under conditions of limited and vague user feedback, and
(4) Ensure low latency to support real-time or near-real-time responses.

\framework builds on the contextual multi-armed bandit framework, with insights from SVMs~\cite{burges1998tutorial} and kNN classifiers~\cite{cover1967nearest}. During an offline pre-processing phase, \framework scatters \textit{signal indicators} across the document embedding space. These signals are obtained from user feedback (e.g., specific documents from past queries) or generated synthetically from document content~\cite{rhyde}.
At inference time, \framework employs a novel mechanism to retrieve a superset of potentially relevant documents.
A two-track ranking algorithm is developed to re-rank the documents based on their relative importance and the collected signals (analogous to bandit weights), and the top-\(k\) most relevant documents are submitted to a LLM for answer generation. \framework maintains flexibility and robustness in document selection by only using these signals to \textit{influence} the ranking as opposed to determining the final selection. By integrating synthetic and real feedback in this manner, \framework offers a universal, scalable, and efficient approach to improve retrieval accuracy.

\subsubsection*{\textbf{Contribution}}
We integrate \framework into \system~\cite{dricopilot}, an internal tool used by \companyname software engineers to assist with various development tasks, including coding, debugging, and documentation, as well as by on-call engineers for incident mitigation. \system serves thousands of users and features a comprehensive tracking system that enables \framework to associate user feedback with specific documents in the knowledge base. Additionally, since \system prompts the LLM to cite its references, \framework effectively links user feedback with the LLM’s downstream preferences, ensuring alignment and improving response quality over a range of commonly encountered scenarios.

\section{Related Work}
\label{sec:background}


\subsection{LLMs as an Assistant}
\label{sec:llm-assistant}
Recent studies in large language models (LLMs) have demonstrated their remarkable capabilities in question answering, text generation, reasoning, and instruction (i.e., prompt) following~\cite{10.5555/3600270.3602281}. These capabilities have opened up opportunities for the deployment of LLMs as technical copilots in various contexts, such as supporting engineers in error diagnosis and troubleshooting~\cite{gonccalves2011collaboration,roy2024exploring}. However, LLMs suffer from several issues that make deployment challenging.

(1) \textbf{Knowledge Usage}: LLMs lack knowledge from proprietary documents available in internal knowledge bases and are prone to make up incorrect answers (i.e., \textit{hallucinate}) when uncertain~\cite{zhang2023sirenssongaiocean}. LLMs can be fine-tuned with data from internal knowledge bases and user feedback~\cite{dai2022promptagatorfewshotdenseretrieval,10.5555/3600270.3602281} to alleviate these issues. However, this process is complex, computationally expensive, and requires periodic retraining (or fine-tuning) as knowledge bases continuously mutate (i.e., documents are created, edited, and deleted over time).

(2) \textbf{Stochasticity and Latency}: Passing a slightly modified prompt to an LLM is likely to result in different textual and semantic outputs~\cite{li2024escapeskyhighcostearlystopping}, more so when the user query is underspecified. This variability makes consistent interactions difficult. Furthermore, state-of-the-art LLMs (e.g., GPT-4o~\cite{gpt4o}) incur inference latencies on the order of seconds to tens of seconds, rendering sampling and multi-round interactions impractical for time-sensitive scenarios.



\subsection{Similarity Cache}
\label{sec:sim-cache}

To address stochasticity and latency issues, prior work has proposed caching~\cite{zhu2024towards,bang2023gptcache}. 
These caches use a \textit{matching function} to determine if a historical response can answer a different query. Matching can be based on textual equality, BERT~\cite{bert}, LLM embeddings~\cite{embedding}, or cosine similarity~\cite{bm25}. While these methods effectively reduce latency and control stochasticity, they are inherently inflexible---struggling to adapt to evolving knowledge bases or leverage user feedback from rephrased or semantically similar queries. Moreover, by focusing only on correct matches, these mechanisms miss insights from errors, limiting the system’s ability to learn and adapt.


\subsection{Retrieval Augmented Generation}
\label{sec:rag-background}



Recent research has proposed retrieval augmented generation (RAG) framework to address knowledge misalignment and hallucinations by providing relevant documents from a knowledge base to the LLM and instructing it to answer the query with those documents.


By recasting as a retrieval problem, RAG lowers the risk of hallucination. However, RAG does not incorporate user feedback and still suffers from misalignment. As RAG lacks user preferences, it instead uses a \textit{proxy metric} (e.g., cosine distance, BM25 score~\cite{bm25}) to evaluate query-document relevance during retrieval. As a result, a highly scored document may actually be semantically irrelevant to the user query due to various factors (e.g., context, related topics). Without the ability to incorporate domain expert feedback, RAG is susceptible to 
generate incorrect answers. 
Recent advancements in RAG systems have introduced dynamic and iterative retrieval strategies~\cite{trivedi-etal-2023-interleaving,shi-etal-2024-generate,10448015,asai2024selfrag, jiang-etal-2023-active}, which enhance relevance but introduce additional latency, making them unsuitable for real-time interactive copilot systems where fast responses are crucial.

\subsection{Hypothetical Query Embeddings (HyQE)}
\label{sec:rhyde}
Hypothetical Query Embeddings~\cite{rhyde} aims to address the misalignment through generating hypothetical queries. Each document chunk is processed by a generator LLM (e.g., GPT-4o) that produces a set of queries for which the chunk contains answers for. During inference, the user’s query is also matched against these \textit{generated queries} rather than just the document chunk. Although this improves recall by fixing query-document misalignment,
ranking documents retrieved from HyQE and other strategies (e.g., full-text, embedding) remains open. In addition, HyQE does not inherently provide mechanisms to incorporate user feedback.

\subsection{Adapting Embeddings}
\label{sec:nudge}
To account for the misalignment between queries and documents, RAFT~\cite{zhang2024raft} proposes retrieval augmented fine-tuning. RAFT fine-tunes the downstream LLM with triplets of queries, answers, and retrieved documents to improve the quality of the generated answer. However, this has high computational overhead and does not address cases where retrieval fails to find the document.

An alternative approach proposed by the \nudge framework~\cite{zeighami2024nudge} performs lightweight embedding adjustment with labeled query-document pairs. Although this approach is simpler than fine-tuning the entire embedding model, it has several drawbacks: (1) It requires training entirely new embeddings upon document or feedback updates. (2) Due to its coarse-grained adjustment to a high-dimensional embedding space, \nudge is prone to impacting unrelated questions, an issue further compounded by the limited quantity of labeled data. (3) \nudge only adapts to positive feedback, whereas \textit{negative} examples (i.e., document \textit{irrelevance}) are more important to learn from and adjust to. Similar to SVMs, focusing on borderline negative examples exposes flaws in the original embedding. Using these instances for fine-tuning enhances the embedding space's ability to discern among potentially confusing documents and provide essential semantic insights.
\section{Indicator Signals}
\label{sec:intuition}

We seek a mechanism to incorporate user feedback and enable the RAG system to evolve its answers over time, subject to two constraints from production deployments. Constraint \#1 is using off-the-shelf models due to the dynamic nature of knowledge bases. Constant document changes make it difficult to keep models up to date and continuous fine-tuning is computationally expensive in terms of resources and operational overhead. Constraint \#2 is minimizing response latency, which reflects the practical requirements of day-to-day working environments. High latency in generating answers (due to frequent LLM calls) would disrupt workflows and reduce the system's utility as a real-time technical assistant, particularly in fast-paced or dynamic settings. We next discuss our mechanism designed to operate effectively within these constraints.

\subsection{Push-Pull Indicators}

We define an \textit{indicator repository}, $\mathcal{I}$, which consists of triplets $(q,d,s)$ where $q$ represents a historical user query, $d$ is a document in the knowledge base, and $s$ is a \textit{signal} indicating how useful $d$ is for answering $q$, ranging from -1 to +1.
%
We observe that historical interactions can serve as valuable guideposts for addressing future queries. For instance, if a feedback triplet $(q_1,d_1,+1)$ exists, then for a subsequent query $q_2$ that is \textit{similar} to $q_1$, the retrieved document list for $q_2$ can be potentially augmented with $d_1$. Conversely, documents associated with a \textit{low} signal can be omitted. The formal definition of query similarity is provided in \cref{sec:arch}.

\begin{figure}[t!]
    \centering

    \includegraphics[width=\linewidth]{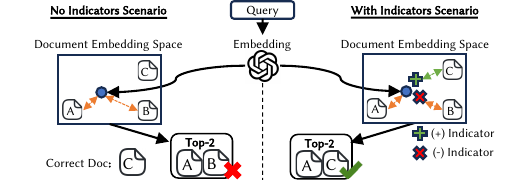}
    \caption{
        \captionTitle{Push-Pull Indicators} Illustrates how indicators influence document retrieval in two scenarios. In \textit{No Indicators}, we incorrectly retrieve \texttt{(A,B)} based on embedding distance. In \textit{With Indicators}, we correctly retrieve \texttt{(A,C)} due to a positive indicator that \textit{pulls} \texttt{Doc-C} closer and a negative indicator that \textit{pushes} \texttt{Doc-B} away.
    }
        \vspace{-0.4cm}
    \label{fig:indicators}
\end{figure}

We provide a running example of this behavior in \cref{fig:indicators}. In \textit{No Indicators}, we set up 
a scenario where misalignment causes the relevant document (\texttt{Doc-C}) to be located far from the query in embedding space, resulting in its exclusion during retrieval. In \textit{With Indicators}, we incorporate feedback indicators, such as $(q_1,D_C,+1)$ and $(q_2,D_B,-1)$ that dynamically alter the retrieved documents. Given a new query $q_3$, which is similar to $q_1$ and $q_2$, \texttt{Doc-C} is \textit{pulled closer} (due to its relevance to $q_1$) and \texttt{Doc-B} is \textit{pushed away} (due to its irrelevance to $q_2$). With these feedback signals, we increase the \textit{likelihood} of retrieving the most relevant documents. In addition, these signals enable the system to adapt by dynamically retrieving different document sets that are combined and holistically re-ranked, resulting in more accurate responses over time.

\subsection{Indicator Generation}
\label{sec:indicator_gen}

The most straightforward mechanism to generate indicators is directly from historical user feedback, referred to as \textit{historical feedback indicators}. However, such feedback is often limited. From internal data usage over a few months, we observe that only 2.4\% (\sqldb) and 4.4\% (\sqldw) have user feedback.

As a result, we augment real user feedback with \textit{synthetic indicators}. We build on top of the technique proposed in HyQE~\cite{rhyde}. From each document in the knowledge base, we prompt an LLM to generate synthetic queries $q$ that can be answered by the document. We ensure, with a high degree of confidence, that each query is related to the document by prompting the LLM to answer its generated query with an answer from the document. In this way, we enrich the limited historical feedback available with a larger bank of synthetic $(q,d,+1)$ triplets mined from the knowledge base.

\subsection{Indicator Granularity}
Due to context window restrictions, documents are split during preprocessing into chunks at logical boundaries (e.g., sections, paragraphs) with a maximum chunk size. Depending on the use-case, indicators can be attached at the \textit{document} or \textit{chunk} granularity. We propose a hybrid approach based on the indicator's source.

\paragraph{Synthetic Indicators:} For generated indicators, we attach indicators at the \textit{chunk} granularity. As LLM generates queries on a per-chunk basis, this naturally aligns a generated query with its source chunk, as opposed to the document as a whole.

\paragraph{Historical Feedback Indicators:} For these indicators, we generalize the indicator and opt for \textit{document} granularity. We adopt a user-centric approach, positing that if the \textit{most relevant} chunk of a document is useful (or not useful), then the document as a whole should be considered similarly useful (or not useful). Although this observation coarsens the indicator, it also provides resilience to structural changes that shift content (e.g., new paragraph).
\section{Architecture}
\label{sec:arch}

\begin{figure}[t!]
    \centering

    \includegraphics[width=\linewidth]{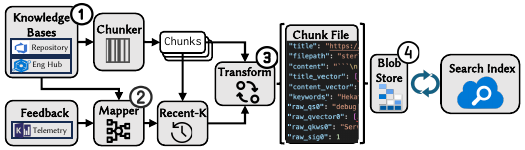}
    \caption{
        \captionTitle{Pre-processing Phase} Overview of the offline pre-processing phase. In this phase, knowledge base documents are processed into a search index.
    }
    \label{fig:arch_preprocess}
        \vspace{-0.5cm}
\end{figure}

We next present the architecture of \framework. Building upon the RAG framework, \framework incorporates user feedback with a two-track ranking mechanism that integrates indicator signals to fairly re-rank documents across multiple retrieval strategies from a single underlying knowledge base. \framework is agnostic to fine-tuning and works with off-the-shelf models, while continuously learning over time to adapt to user preferences, ensuring scalability and efficiency.

\subsection{Preprocessing Phase}\label{sec:preprocessing}

The preprocessing phase of \framework, illustrated in \cref{fig:arch_preprocess}, adopts a \textbf{document-centric} approach for maintaining and processing information.
\circled{1} \framework periodically retrieves the latest user feedback from internal telemetry along with the corresponding documents referenced during those conversations and \circled{2} builds a mapping from document to available user feedback, trimmed to the most recent $K$ entries per document.
\framework then chunks each document. \circled{3} Each document chunk is passed through a transform step that extracts the required fields (e.g., title, content, keywords), generates embeddings with a model (e.g., \texttt{text-embedding-3-large}~\cite{embedding}), synthesizes hypothetical questions for generating synthetic indicators (\cref{sec:rhyde}), and attaches any relevant historical feedback indicators. For each indicator, we store the original query, the query's embedding, the query keywords, and its signal value (i.e., document's usefulness to the query, ranging from $-1$ to $+1$).
\circled{4} \framework writes all generated fields to remote cloud storage, and the artifacts are periodically incorporated into a search index. This process enables \framework to efficiently capture document and feedback updates in the knowledge base and incorporate them into the search index for online retrieval, ensuring the system remains up to date.

This preprocessing step circumvents the need for any training process to modify the embedding, as required in~\cite{zeighami2024nudge,zhang2024raft}. With more feedback collected, we simply ``\textit{append}'' more indicators to the relevant document chunk(s), which will be used in aggregate during retrieval (discussed in the next section). 
The retrieval process can seamlessly incorporate indicator information as metadata alongside the retrieval results from other strategies such as~\cite{knn,karpukhin-etal-2020-dense}. 

\subsection{Online Retrieval Phase}
\label{sec:arch-online}

\begin{figure}[t!]
    \centering

    \begin{subfigure}{\linewidth}
        \includegraphics[width=\linewidth]{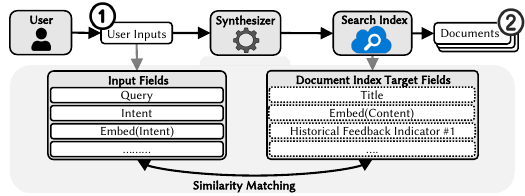}
        \caption{Retrieve}
        \label{fig:arch_retrieve}
    \end{subfigure}

    \begin{subfigure}{\linewidth}
        \includegraphics[width=\linewidth]{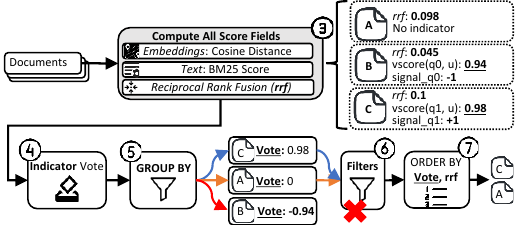}
        \caption{Score and Re-Rank}
        \label{fig:arch_rerank}
    \end{subfigure}
    \vspace{-0.5cm}
    \caption{
        \captionTitle{Inference Phase} Overview of the online inference phase. In this phase, \framework retrieves a superset of documents, re-ranks them, and outputs the most relevant documents.
    }
    \label{fig:arch_online}
    \vspace{-0.5cm}
\end{figure}

We next discuss the online retrieval phase, illustrated in \cref{fig:arch_online}. This phase is divided into three separate subphases: (1) retrieve (\cref{fig:arch_retrieve}), (2) two-track scoring, and (3) two-track re-rank (\cref{fig:arch_rerank}). We first discuss the retrieval subphase.

\subsubsection*{\textbf{\textit{Retrieve}}}
As shown in \cref{fig:arch_retrieve}, the user provides \circled{1} multiple queries to \framework. These inputs can be the user's original query or alternative rewritten queries for context retrieval~\cite{li2024dmqrragdiversemultiqueryrewriting}.
The search index then retrieves the \(M\) most relevant documents based on how \textit{similar} the input fields are to search index fields. However, there are a combinatorial number of matchings between input fields (e.g., query, intent embedding) and search index fields from the document chunk (e.g., title, content embedding). Furthermore, similarity can be defined across multiple relationships (e.g., query$\sim$title and query embedding $\sim$ content embedding). We present a detailed discussion on selecting retrieval strategies in \cref{sec:eval}. At present, \framework requires the user to specify the set of retrieval strategies.

Based on user specifications, \circled{2} \framework synthesizes and executes all retrieval strategies, combining each strategy's resulting document set. Let $L_s$ represent the chunks retrieved by strategy $s$. This phase produces the union of all sets across strategies, denoted as $L = L_1 \cup \dots \cup L_s~\forall s$. \framework then transitions to the next phase, where it selects the most relevant ones from $L$. Note that any search strategy using indicator fields (e.g., query, keywords) retrieves documentation referenced in previously similar queries, thereby augmenting the retrieved documentation set.

\subsubsection*{\textbf{\textit{Two-Track Scoring}}}
The scoring subphase \circled{3} computes two different score classes for each chunk: \textbf{relevance scores} that quantify similarity based on user input and document fields and \textbf{vote scores} derived from indicators (see \cref{sec:indicator_gen}). We discuss this next.

\textit{Relevance Scores} measure the similarity between user inputs and document chunks using various document fields. We employ cosine for embedding-based fields, denoted as \verb|vscore|:
{\footnotesize
\begin{align}
    \text{vscore}\left(e(u),e(v)\right) = \frac{1}{2 - \frac{\langle e(u), e(v) \rangle}{\left\lVert e(u) \right\rVert \left\lVert e(v) \right\rVert}},
    \label{eq:vscore}
\end{align}}
where \(e(u)\) represents the embedding of the user's query $u$, and \(e(v)\) represents the embedding of a document chunk field $v$, such as its title or content.
For text-based fields, we utilize the BM25~\cite{bm25} score and denote it as \(\text{tscore}(u, v)\). By considering multiple fields such as content, title, content embeddings, and title embeddings, several similarity scores can be calculated for each chunk. When these scores are used in the retrieve phase discussed earlier, they are typically included in the returned ``similarity score'' along with the retrieved set of documents, computed by the search algorithm.

\framework then combines these similarity scores across all fields into a single aggregated score using Reciprocal Rank Fusion (\verb|rrf|)~\cite{rrfpaper}. This algorithm assigns higher scores to chunks that consistently rank well across multiple ranked lists. The \verb|rrf| score for a document chunk \(d\) among the list of chunks \(L\) is defined as:
{\footnotesize
\begin{align}
    \text{rrf}(d, L) &= \sum_{L^{(f)} \in L} \frac{1}{60 + rank\left(d, L^{(f)}\right)}.
    \label{eq:rrfscore}
\end{align}
}
Each list $L^{(f)}$ is derived from ordering all retrieved document chunks \(L\) by a specific field $f$, such as a \verb|vscore| or \verb|tscore|. Chunks ranked higher in more lists receive a higher \verb|rrf| score. Thus, we combine multiple disparate scores into a single robust score.

\textit{Vote Scores} quantify the impact of feedback indicators on a chunk’s relevance to the current user's query $u$. 
For a document chunk \(d\), the set of relevant indicators is denoted as \(\mathcal{I}_d\), where each indicator, indexed by $i$, is represented as a triplet \((q_i, d, s_i) \in \mathcal{I}_d\). Here, \(q_i\) represents the query (e.g., historical user query) recorded for this indicator. The signal \(s_i\) from \(-1\) to \(+1\) indicates if \(d\) provided good answers for \(q_i\). The intuition is that if the recorded historic query \(q_i\) of this indicator is very similar to the current user query \(u\), then this indicator should be very relevant. Moreover, if the signal value $s_i$ is high, indicating that this document answers query \(q_i\) well, it should also answer \(u\) relatively well, thus a higher ``vote''. The vote score for document \(d\) is computed as follows:
{\small
\begin{align}
    c(u, q_i) &= 
    \begin{cases} 
        \text{vscore}(e(u), e(q_i)) & \text{if } \text{vscore}(e(u), e(q_i)) \geq T, \\ 
        0 & \text{otherwise}
    \end{cases} \label{eq:vote-sig} \\
    \text{vote}(d) &= \frac{\sum_{(q_i, d, s_i) \in \mathcal{I}_d} \sum_{u \in U} (c(u, q_i) \cdot s_i)}{\sum_{(q_i, d, s_i) \in \mathcal{I}_d} \sum_{u \in U} \mathbbm{1}{(\text{vscore}(e(u), e(q_i)) \geq T)}},
    \label{eq:vote}
\end{align}
}
where,

\begin{tabular}{lp{6.9cm}}
$U$: & user inputs (e.g., $\{\text{query, rewritten intent}\}$), \\
$T$: & the \threshold level to consider an indicator relevant. \\
\end{tabular}


\verb|vote| aggregates and normalizes the weighted signal strengths of all indicators based on the similarity between its recorded query $q_i$ and the current query $u$. This ensures that the score reflects only the weighted contribution of relevant indicators. Any indicators with \verb|vscore| lower than $T$ are ignored. Although this induces a nonlinearity by sharply pruning any irrelevant indicators, this allows us to utilize a 0 vote for chunks that lack relevant indicators.

\subsubsection*{\textbf{\textit{Two-Track Re-Rank}}}
\cref{fig:arch_rerank} shows how the previously computed scores are used for re-ranking.
In \circled{5},
\framework groups all chunks by their respective votes and aggressively prunes any vote-groups with a negative vote, as a negative score indicates that the associated chunks are not useful for answering the current query. In the example shown, \verb|Doc-B| is pruned because it has a vote score of -0.94. This pruning enables effective learning from previous retrieval errors.

\begin{outputfull}
\circled{6} Within each vote group, \framework applies two optional filters: (1) \textit{k-truncation} and (2) \textit{margin filter}. The k-truncation filter selects the top-$k$ most relevant chunks from each vote-group based on their \verb|rrf| scores (\cref{eq:rrfscore}). The margin filter removes all chunks with an \verb|rrf| score that is below a tunable percentage of the \textit{most relevant chunk}'s \verb|rrf| score. These filters ensure a balanced selection of chunks:
\begin{itemize}
    \item 
    The k-truncation filter 
    prevents large documents with many chunks from disproportionately dominating the output solely due to a single strong feedback indicator. By limiting the number of chunks per vote-group, this filter ensures that other relevant documents and their associated chunks are also considered for downstream processing, thereby maintaining ``diversity'' in the final retrieved results.
    \item The margin filter ensures that only highly relevant chunks, which are sufficiently close in relevance to the top chunk, are passed to the downstream LLM. By discarding less relevant chunks, this filter enhances the efficiency and accuracy of the final response generation.
\end{itemize}
\end{outputfull}

\begin{outputabridge}
As an LLM's performance may be negatively impacted due to superfluous documents~\cite{chen2023benchmarkinglargelanguagemodels}, \circled{6} \framework applies two additional filters within each vote group to ensure a balanced selection of chunks. (1) \framework first selects the top-$N$ most relevant chunks within each vote-group based on \texttt{rrf} scores~(\cref{eq:rrfscore}). This prevents a large document with many chunks from overflowing the output document set. (2) \framework then uses a margin filter to remove all chunks with a \texttt{rrf} score below a user-specified percentage of the \textit{most relevant chunk}'s \texttt{rrf} score. This margin filter has been shown to effectively reduce context length when a dominant relevant chunk is present~\cite{dricopilot}.
\end{outputabridge}

\circled{7} \framework then globally orders all remaining chunks first by \verb|vote| scores~(\cref{eq:vote}) and then by \verb|rrf| scores~(\cref{eq:rrfscore}). This \textit{two-track ranking} system prioritizes chunks with more positive feedback indicators (i.e., fast-tracked) while refining the order of chunks within the same vote group based on relevance. However, negative vote pruning may reduce (or eliminate) the chunks that \framework retrieves. To address this, \framework repeats \circled{1}-\circled{7} until the number of final chunks exceeds a target percentage of the requested top-$k$. In each iteration, \framework increases the number of chunks retrieved by its retrieval strategies.
Experiments show that iteration numbers remain low. Moreover, since no additional LLM call is made, each iteration can be done quickly and thus does not add much latency.

\subsubsection*{\textbf{\textit{Discussion}}}
The two-track ranking system prioritizes the indicator signal over the \verb|rrf| score, effectively promoting previously relevant documents to the top of the ranking and demoting irrelevant ones. 
With a filter $N<<k$, we avoid overflowing the documents being promoted based on indicator signals, making the algorithm more robust to embedding misalignment. 
Moreover, the two-track ranking system is adaptable; the \verb|rrf| component can be replaced with any state-of-the-art algorithm (e.g., ~\cite{jiang-etal-2023-active,gupta2024comprehensivesurveyretrievalaugmentedgeneration,ye2024r2agincorporatingretrievalinformation,joycepaper,chang2024communitykgragleveragingcommunitystructures,sarthi2024raptor}).




\begin{outputfull}
    \input{feedback/ConvergeProof}
\end{outputfull}
\begin{outputfull}

\section{Implementation}
\label{sec:impl}
\yz{maybe we can hide this detail if space is really limited for SIGIR? For VLDB, this might be helpful?}

We next discuss the relevant implementation details of \framework, along with insights and guidance
into the parameters exposed by \framework that guide retrieval and re-ranking.

\subsection{Document Search Index}
\framework supports populating any document search indexes (e.g., Azure Cognitive Search). 
\framework requires the provided search index to support indexing multiple text and embedding
fields. In addition, the search index must support \textit{hybrid search} \sk{I didn't see the experiments which the improved effectiveness of hybrid search}. Hybrid search allows
finding documents that score highly based on a combination of similarity conditions (e.g.,
question embedding $\sim$ title embedding \verb|AND| question $\sim$ content). Hybrid search enables
\framework to directly acquire rankings based on global corpus scores. Although \framework may
benefit from semantic search capabilities, semantic search is not required.

\subsection{Number of Indicators}
Storing more targeted indicators, whether synthetic or historical, generally enhances retrieval, as it enables a larger corpus to be fetched prior to refinement and curation. However,
increasing the number of indicators also results in increased space usage and greater latency during online inference.
By default, \framework only generates 5 synthetic questions, due to preprocessing overhead
(e.g., increased LLM overhead) and online inference overhead (e.g., building Lucene index,
additional retrieval strategies). Nevertheless, we observe that larger or more content-rich chunks
could benefit from more synthetic questions for improved coverage. We leave dynamically determining
an adequate number of synthetic indicators based on chunk content for future work.

For historical user feedback, we frame it as a classical \textit{caching problem}, where the most relevant and useful feedback must be selected and stored as indicators for each document. By default, \framework first de-duplicates and eliminates any indicators
that are no longer valid (e.g., when a document has been updated since feedback was given). \framework then
retains the most recent $K$ indicators per document. We leave alternative approaches, such as clustering
indicators or generalizing indicators across document changes, to future work.

\subsection{Indicator Signal Mapping}
\label{sec:impl-indicator-mapping}

By default, \framework assumes feedback is given on a standard 1-5 star rating.
From this, \framework makes two broad generalizations: (1) a high rating implies the retrieved
documents are relevant, and (2) a low rating implies the retrieved documents are irrelevant.
With this, \framework linearly maps the 1-5 star rating to \verb|[-1, 1]|.

In the event that the output contains a reference list, \framework further refines each document's
score with its ranking in the reference list. For instance, given a star rating of 5 (e.g.,
documents are relevant) and a document list \verb|Doc-A,B,C|, \framework will assign indicator
signals of 1.0, 0.75, and 0.5, respectively. We emphasize that \framework natively supports
arbitrary feedback schemes, so long as they are mappable to a continuous scale from
\verb|-1| (highly irrelevant) to \verb|1| (highly relevant), with \verb|0| indicating neutral
or unknown document relevance.

As the standard 1-5 star rating is coarse, the generalizations \framework adopts may be prone
to error. Specifically, the LLM-generated answer based on the documents may have a low user
rating for other reasons (e.g., user misclicked, generated code is invalid). Synergistically,
as the rating allows for more targeted feedback on specific documents, \framework will naturally
generate more specific and useful indicators.

\subsection{Re-rank Hyperparameters}

The re-rank phase (see \cref{sec:arch-online}) defines four parameters: (1) \textit{top-$k$} that
controls how many documents \framework should output, (2) \textit{threshold} that controls when an
indicator should be incorporated into the vote, (3) \textit{vote-group $k$-truncation} that
controls how many chunks to preserve per vote-group, and (4) margin filter's truncation percentage for filtering low
relevance chunks.

Selecting these parameters requires hyperparameter optimization. \framework selects
these parameters by performing a grid search against the target knowledge base using a curated
validation set of historical user questions and their retrieved documents. For all experiments, we
provide a sensitivity analysis over different values of \textit{top-$k$} or \textit{thresholds}.
Based on an internal validation set, we use a \textit{vote-group $k$-truncation} of 1 and no
margin filter.

\end{outputfull}

\section{Evaluation}
\label{sec:eval}
\begin{outputfull}
We implement the preprocessing phase of \framework using periodic cloud-based machine learning pipelines to process documents from internal knowledge bases (e.g., Wiki pages, Git repositories) into JSON artifacts that are then indexed by a cloud-based search service~\cite{azureai}. We integrate \framework's inference phase into \system's document retrieval pipeline, enabling \system to incorporate feedback. For embeddings, we utilize \texttt{\embedding}~\cite{embedding} for generating embeddings and \chat for prompting.
\end{outputfull}

We evaluate \framework through its integration with \system.
\system provides \framework with both the user query and a rewritten intent. We configure \framework to generate 5 synthetic indicators for each chunk during preprocessing. We initialize with all available historical feedback and allow \framework to maintain up to 6 (\sqldb) and 18 (\sqldw) user feedback indicators, stored in descending time order to prioritize more recent feedback. As we primarily evaluate \framework's retrieval performance, we disable its margin filter and select at most $N=1$ chunk from each vote-group to favor more diverse documents. 

\begin{outputfull}
\subsection{Retrieval Strategies}
We first discuss the retrieval strategies utilized by \framework throughout all experiments. The underlying index search service supports hybrid search queries by running each constituent similarity search (e.g., user query against title, user query embedding against content embedding) in parallel. It then combines the chunks retrieved by each constituent probe into a global order using reciprocal rank fusion (\cref{eq:rrfscore}).

As \framework relies on this index search service for first-stage retrieval and re-ranking, it can utilize various retrieval strategies. Here, we distinguish between \textit{indicator} fields and \textit{dense} fields. \textit{Dense} fields are those that exist for all chunks and provide substantive information about the chunk (e.g., title, content, entities, keywords). 
\textit{Indicator} fields, on the other hand, are derived from feedback indicators. These fields contain information such as the prior or synthetic querys associated with a chunk, the embeddings of those querys, and their corresponding feedback signals from the indicators.

A strawman solution would be to perform a single hybrid search using all fields together, i.e., all the indicator fields and dense fields.
However, this approach is ineffective due to the search service's use of rank fusion. 
Since indicators correspond to prior or synthetic querys, they are unlikely to ALL strongly match an incoming user query.
With more indicators being combined into the query, a document with only one strongly matching indicator may have its relevance diluted by the remaining low-relevance indicators, resulting in the document not being retrieved.

To address this, we configure \framework to execute multiple parallel searches. \framework performs
one hybrid search using the available \textit{dense} fields. This approach is effective, as a user query that strongly matches multiple dense fields (i.e., fields describing the chunk's content) indicates higher relevance. 
We then perform individual searches against
each indicator to retrieve chunks that have a strong similarity match (for at least one indicator). This configuration enables
\framework to retrieve a superset of what is needed, before re-ranking to obtain the most relevant
documents.

\yz{top M? or top-K? or top-k, might need a thorough check}
We configure \framework to retrieve the top-$k$ chunks for hybrid search. We allow \framework to 
initially retrieve up to 100 chunks for pure vector search or 50 chunks for hybrid search against each
indicator. \framework computes its local \verb|rrf| (\cref{eq:rrfscore}) over the
available dense fields from the retrieved documents. We allow \framework to incrementally expand the
retrieval size (i.e., double) until \framework retrieves at least half of $k$ chunks.
A sensitivity analysis of the specific dense fields used is presented in \cref{sec:eval-exact-high}.
\end{outputfull}

\subsection{Metrics}

\subsubsection*{Recall}
\label{sec:recall}
Rather than relying on LLM-based metrics~\cite{10.3115/1073083.1073135, NEURIPS2021_e4d2b6e6, METEOR, sellam-etal-2020-bleurt}, which often lack the domain-specific insight needed for accurate evaluation~\cite{dricopilot}, we adopt standard recall metric~\cite{bishop2006pattern} and de-prioritize precision due to larger LLM context windows.
Using telemetry from \system, we extract only ``referenced'' documents from the LLM's output to define the ``golden'' set and compute recall as:
{\footnotesize
\begin{equation}
    \text{recall} = \frac{\text{len}(\mathtt{golden} \cap \mathtt{retrieved})}{\text{len}(\mathtt{golden})}.
    \label{eq:recall}
\end{equation}
}

\subsubsection*{Hit@N}
We use \textit{Hit@}$N$ to assess whether at least one relevant item appears in the top-$N$ retrieved results. This is effective when a single correct document suffices for optimal performance, such as for the synthetic queries. Formally:
{\footnotesize
\begin{equation}
    \text{Hit@}N = 
    \begin{cases}
        1 & \text{if } \mathtt{golden} \cap \mathtt{retrieved}_{[:N]} \neq \emptyset \\
        0 & \text{otherwise}
    \end{cases}
    \label{eq:hitn}
\end{equation}
}
Here, $\mathtt{golden}$ is the ground truth, and $\mathtt{retrieved}_{[:N]}$ the top-$N$ results. Averaging over queries yields the final Hit@{$N$} score.

\begin{outputfull}
We do not generate an LLM answer or extract its reference list from the newly retrieved set of documents. This decision is guided by two main factors: (1) the inherent stochasticity and variation in LLM-generated answers, which we cannot reliably control, and (2) potential truncation of LLM answers due to token limits. Instead, we evaluate the ``golden'' reference list directly against the newly retrieved set of top-$k$ documents.

\subsection{Lucene Overhead}
Before evaluating \framework in different scenarios, we conduct a microbenchmark study to assess the overhead of incorporating Lucene~\cite{X}, an open-source Apache project for full-text indexing, scoring, and search. Using a compute instance with 2vCPUs and 8G memory, we process approximately 20,000 document chunks from \sqldb and evaluate single-threaded build and scoring times across different fields. Each configuration is executed five times, and the mean values are reported.

\begin{table}
    {\footnotesize\begin{tabular}{c|c|c|c|c|c}
    \toprule
    \multicolumn{1}{c}{\textbf{\# Chunks}} \vline
    & \multicolumn{1}{c}{\textbf{Base}} \vline
    & \multicolumn{1}{c}{\textbf{Base,Feedback}} \vline
    & \multicolumn{1}{c}{\textbf{Base,RHyDE}} \vline
    & \multicolumn{1}{c}{\textbf{All}} \\
    \midrule

    100 & 4s & 4s & 6s & 6s \\
    1000 & 4s & 4s & 5s & 7s \\
    10000 & 4s & 4s & 5s & 6s \\
    All & 4s & 4s & 5s & 6s \\
    \bottomrule
\end{tabular}
}
    \caption{
        \captionTitle{Lucene Index Build} Presents the build times of constructing a Lucene index
        over varying numbers of chunks and document fields. ``Base'' builds a Lucene index
        over the chunk title and chunk content. Each evaluation presents the mean of 5 separate
        single-threaded invocations.
    }
    \label{tab:lucene}
\end{table}

\begin{table}
    {\footnotesize\begin{tabular}{c|c|c|c|c|c}
    \toprule
    \multicolumn{1}{c}{\textbf{\# Chunks}} \vline
    & \multicolumn{1}{c}{\textbf{Base}} \vline
    & \multicolumn{1}{c}{\textbf{Base,Feedback}} \vline
    & \multicolumn{1}{c}{\textbf{Base,RHyDE}} \vline
    & \multicolumn{1}{c}{\textbf{All}} \\
    \midrule

    100 & 1s & 1s & 1s & 1s \\
    1000 & 1s & 3s & 1s & 4s \\
    10000 & 6s & 28s & 11s & 38s \\
    All & 13s & 58s & 22s & 80s \\
    \bottomrule
\end{tabular}
}
    \caption{
        \captionTitle{Lucene Index Score} Presents the scoring times over varying numbers of
        chunks and document fields. ``Base'' scores over the chunk title and chunk content. Each
        evaluation presents the mean of 5 separate single-threaded invocations.
    }
    \label{tab:lucene-read}
\end{table}

The build results are presented in \cref{tab:lucene}, and the scoring results in \cref{tab:lucene-read}. From \cref{tab:lucene}, we observe that index build times remain low, typically between 4--7 seconds. Moving to scoring times in \cref{tab:lucene-read}, we note that incorporating feedback indicators noticeably increases scoring times. This increase is primarily due to the additional Lucene index lookups and score computations required for each full-text feedback indicator field.
We still broadly observe that scoring hundreds of chunks remains within a few seconds, which is
an acceptable overhead. Furthermore, we can speed up scoring by parallelizing it across
available hardware resources.
\end{outputfull}

\subsection{Benchmark: Iterative Learning}

To evaluate the long-term impact of \framework under continuous feedback, we simulate a production environment using \sqldb's document set. Synthetic queries are generated from technical documents via a curated LLM prompt, allowing us to treat the source document as the ``golden'' retrieval target, effectively yielding a labeled test set. Each iteration follows these steps:

\begin{enumerate}
    \item Retrieve documents for each synthetic query.
    \item Generate synthetic feedback based on the retrieved results.
    \item Update the feedback indicator set per Section~\ref{sec:preprocessing}.
\end{enumerate}

Each iteration processes 30 queries: 10 new and 20 repeated from earlier rounds, mimicking real-world query patterns. A 5-star rating is given if the retrieved set includes the ground-truth document; otherwise, a 1-star rating is assigned. This process runs for 200 iterations, totaling 6,000 queries.

\begin{figure}[t]
    \centering
    \includegraphics[width=0.995\linewidth]{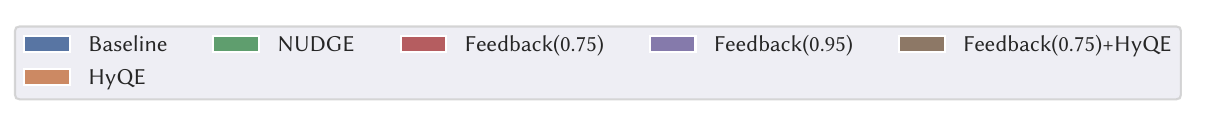}
    \begin{subfigure}[b]{0.45\linewidth}
        \centering
        \includegraphics[width=\linewidth]{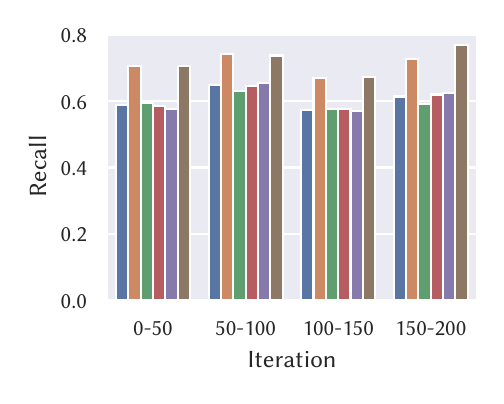}
        \vspace{-0.7cm}
        \caption{Unseen (New) queries}
        \label{fig:eval-overtime-new}
    \end{subfigure}%
    \begin{subfigure}[b]{0.45\linewidth}
        \centering
        \includegraphics[width=\linewidth]{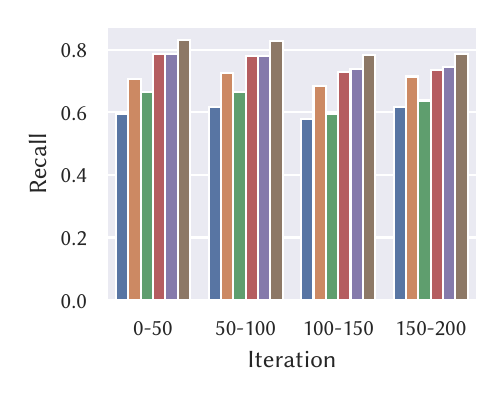}
        \vspace{-0.7cm}
        \caption{Repetitive (Old) queries}
        \label{fig:eval-overtime-old}
    \end{subfigure}
    \vspace{-0.4cm}
    \caption{
        Recall with the state-of-the-art algorithms~\cite{zeighami2024nudge,rhyde}.}
    \label{fig:eval-iterative-learning-bar}
    \vspace{-0.3cm}
\end{figure}

Figure~\ref{fig:eval-iterative-learning-bar} compares the average recall across various configurations: the \textbf{Baseline} configuration, \textbf{NUDGE}~\cite{zeighami2024nudge}, \textbf{HyQE} (synthetic indicators only), \textbf{Feedback} (user indicators only) configurations with thresholds of 0.75 and 0.95, and \textbf{Feedback+HyQE} jointly, which combines indicators from user feedback queries and synthetic queries at a threshold of 0.75. In the \textbf{NUDGE} configuration, positive feedback is used to fine-tune an embedding adjustment function after each iteration, thereby aligning the embeddings of the query and the corresponding document more closely. For the \textbf{HyQE} implementation, we leverage our two-track re-ranking system (as discussed in Section~\ref{sec:arch-online}) and only incorporate votes from LLM-generated queries as synthetic query signals.

For unseen queries (\cref{fig:eval-overtime-new}), recall improvements from indicator-based retrieval (\textbf{Feedback(0.75)}, \textbf{Feedback(0.95)}) and \textbf{NUDGE} are relatively small. In contrast, \textbf{HyQE} and \textbf{Feedback(0.75)+HyQE} show substantial gains, indicating effectiveness of synthetic signals. For repetitive queries (\cref{fig:eval-overtime-old}), \textbf{NUDGE}, \textbf{Feedback(0.75)}, and \textbf{Feedback(0.95)} achieve 6\%, 18\%, and 18\% improvements over the \textbf{Baseline}, respectively. \textbf{Feedback(0.75)+HyQE} outperforms \textbf{HyQE} alone for unseen (10\%) and repetitive (22\%) queries with additional user feedback signals.

\begin{figure}[t!]
    \centering
    \begin{subfigure}{0.3\linewidth}
        \includegraphics[width=\linewidth]{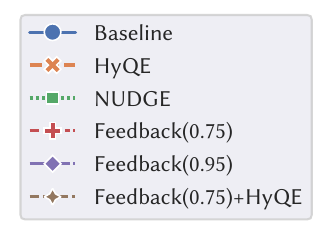}
        \vspace{1.8em}
    \end{subfigure}%
    \begin{subfigure}{0.6\linewidth}
        \includegraphics[width=\linewidth]{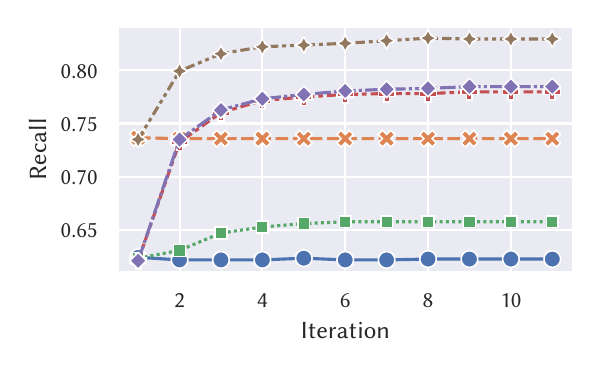}
    \end{subfigure}
    \vspace{-0.5cm}
    \caption{
        Recall for repetitive queries across iterations.
    }
    \vspace{-0.2cm}
    \label{fig:eval-iterative-learning-line}
\end{figure}


Table~\ref{tab:recall_hitn_settings} summarizes the recall and Hit@N performance across all 200 iterations. Both \textbf{HyQE} and \textbf{Feedback(0.75)+HyQE} consistently enhance accuracy for both unseen and repetitive queries. Notably, high-threshold feedback learning (\textbf{Feedback(0.95)}) also slightly improves recall on unseen queries compared to the \textbf{Baseline}, suggesting that user feedback can generalize to new cases. For repetitive queries, \textbf{NUDGE}, \textbf{Feedback(0.75)}, and \textbf{Feedback(0.95)} significantly outperform the \textbf{Baseline}. The combination of \textbf{Feedback(0.75)} with \textbf{HyQE} yields the highest gains, surpassing \textbf{HyQE} alone.
Across all query types, \textbf{Feedback(0.75)+HyQE} demonstrates an 8\% improvement in recall and a 7\% improvement in Hit@10 over \textbf{HyQE}. Unlike \textbf{NUDGE}, our approach integrates multiple reranking signals, achieving optimal overall performance.

\begin{table}[t!]
\centering
\resizebox{\columnwidth}{!}{%
\begin{tabular}{l|cc|cccccccc}
\toprule
\textbf{Setting} 
& \multicolumn{2}{c|}{\textbf{Recall}} 
& \multicolumn{8}{c}{\textbf{Hit@N}} \\
\cmidrule(lr){2-3} \cmidrule(lr){4-11}
 & Old & New 
 & \multicolumn{2}{c}{@3} 
 & \multicolumn{2}{c}{@5} 
 & \multicolumn{2}{c}{@7} 
 & \multicolumn{2}{c}{@10} \\
\cmidrule(lr){4-5} \cmidrule(lr){6-7} \cmidrule(lr){8-9} \cmidrule(lr){10-11}
 & & 
 & Old & New 
 & Old & New 
 & Old & New 
 & Old & New \\
\midrule
\textbf{Baseline} & 0.6055 & 0.6099 & 0.4111 & 0.4025 & 0.4910 & 0.4797 & 0.5296 & 0.5272 & 0.5741 & 0.5777 \\
\textbf{HyQE} & 0.7150 & 0.7179 & 0.5839 & 0.5955 & 0.6344 & 0.6446 & 0.6658 & 0.6688 & 0.6967 & 0.6975 \\
\textbf{NUDGE} & 0.6640 & 0.6468 & 0.5641 & 0.4673 & 0.5950 & 0.5208 & 0.6106 & 0.5490 & 0.6312 & 0.5827 \\
\textbf{Feedback(0.75)} & 0.7830 & 0.7258 & 0.7269 & 0.3980 & 0.7357 & 0.4807 & 0.7412 & 0.5272 & 0.7520 & 0.5757 \\
\textbf{Feedback(0.95)} & 0.7835 & 0.7265 & 0.7302 & 0.4050 & 0.7389 & 0.4861 & 0.7462 & 0.5332 & 0.7570 & 0.5752 \\
\textbf{Feedback(0.75)+HyQE} & \textbf{0.8290} & \textbf{0.7924} & \textbf{0.7853} & \textbf{0.6010} & \textbf{0.7924} & \textbf{0.6505} & \textbf{0.7992} & \textbf{0.6758} & \textbf{0.8065} & \textbf{0.7046} \\
\bottomrule
\end{tabular}
}
\caption{Comparison of Recall and Hit@N.}
\label{tab:recall_hitn_settings}
\vspace{-0.6cm}
\end{table}

Figure~\ref{fig:eval-iterative-learning-line} illustrates recall trends for repetitive queries across iterations. \textbf{NUDGE} steadily improves recall by refining the embedding adjustment function based on positive feedback, yielding a 3\% gain over the \textbf{Baseline}. In contrast, \textbf{Feedback(*)} configurations achieve greater gains (16--22\%) by leveraging positive feedback and re-ranking with negative feedback. \textbf{HyQE} improves recall from iteration 0, while combining it with feedback learning (\textbf{Feedback(0.75)+HyQE}) achieves the best performance by effectively integrating a variety of reranking signals.  



\subsection{Fine-grained Scenarios}
In this section, we evaluate the effectiveness of the algorithms on real user queries based on historic data from two \system deployments for \sqldb and \sqldw (see \cref{tab:motivation}).

\subsubsection*{Exact High-Feedback}
\label{sec:eval-exact-high}

This scenario considers a user issuing a query that \system had previously answered correctly. With a feedback mechanism, the ``golden'' document is expected to be reliably retrieved. We evaluate historical queries where \system performed well, 23 for \sqldb and 103 for \sqldw.

\circled{1} \textbf{Baseline} retrieves documents using an \verb|rrf| score across indexed fields: title, content, and their embeddings, as defined in \cref{eq:rrfscore}.
\circled{2} \textbf{+HyQE}~\cite{rhyde} incorporates synthetic indicators and computes a \verb|vote| score for re-ranking.
\circled{3} \textbf{+Feedback} enhances \textbf{Baseline} by performing additional vector searches using embeddings from user feedback indicators.
Finally, \circled{4} \textbf{+HyQE+Feedback} combines feedback-based and synthetic indicator-based signals.

\begin{table}[t!]
    \centering
    \resizebox{0.8\width}{!}{
    {\small\begin{tabular}{lcccccc}
    \toprule
    & \multicolumn{3}{c}{\textbf{\sqldb Recall@k}} & \multicolumn{3}{c}{\textbf{\sqldw Recall@k}} \\
    \cmidrule(lr){2-4} \cmidrule(lr){5-7}
    & $3$ & $7$ & $12$ & $3$ & $7$ & $12$ \\
    \midrule
        \circled{1} \textbf{Baseline} & 0.62 & 0.79 & 0.79 & 0.56 & 0.72 & 0.76 \\
        \circled{2} \textbf{+HyQE} & 0.50 & 0.71 & 0.79 & 0.57 & 0.72 & 0.75 \\
        \circled{3} \textbf{+Feedback} & \textbf{1.00} & \textbf{1.00} & \textbf{1.00} & \textbf{0.99} & \textbf{0.99} & \textbf{0.99} \\
        \circled{4} \textbf{+HyQE+Feedback} & \textbf{1.00} & \textbf{1.00} & \textbf{1.00} & \textbf{0.99} & \textbf{0.99} & \textbf{0.99} \\
    \bottomrule
\end{tabular}
}}    
    \caption{
        \captionTitle{Exact High-Feedback} ``Golden'' document is consistently retrieved when highly correlated
        feedback is present.
    }
    \label{tab:eval-exact-high}
    \vspace{-0.5cm}
\end{table}

We report mean recall at various top-$k$ values in \cref{tab:eval-exact-high}.
\textbf{+HyQE} yields only marginal gains over \textbf{Baseline}, as the added fields introduce slight perturbations without significantly changing \system's relevance evaluation.
In contrast, \textbf{+Feedback} configurations (\circled{6}, \circled{7}) achieve perfect recall across both datasets—unsurprising given the evaluation uses previously successful historical queries.

\subsubsection*{Exact Low-Feedback}

This scenario considers a user issuing a query that \system previously answered poorly. We consider only historical interactions rated 1 or 2 stars, yielding 48 queries for \sqldb and 180 for \sqldw. As in \cref{sec:eval-exact-high}, we compare the \textbf{Baseline} and \textbf{Feedback} configurations, using thresholds $T \in \{0.75, 0.85, 0.95\}$ as defined in \cref{eq:vote-sig}.

Since the correct documents are unknown for this set of low-rated queries, we assess similarity instead of recall. This metric, computed similarly to \cref{eq:recall}, captures the overlap between the historically retrieved and newly retrieved documents—lower values indicate greater divergence.
Across all thresholds the \textbf{Feedback} configuration consistently yields low similarity (close to 0), demonstrating \framework's ability to help \system avoid previously incorrect documents and explore alternative retrievals.

\begin{outputfull}
    \begin{figure}[t!]
    \centering
    \includegraphics[width=\linewidth]{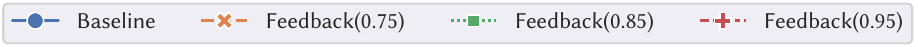}
    \begin{subfigure}{0.5\linewidth}
        \includegraphics[width=\linewidth]{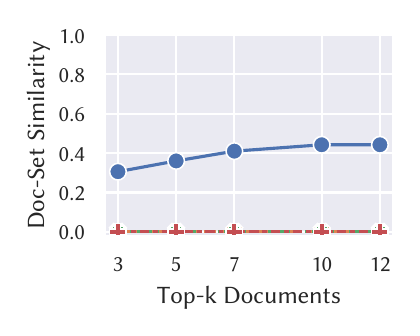}
        \vspace{-0.7cm}
        \caption{\sqldb}
        \label{fig:eval-exact-low-sqldb}
    \end{subfigure}%
    \begin{subfigure}{0.5\linewidth}
        \includegraphics[width=\linewidth]{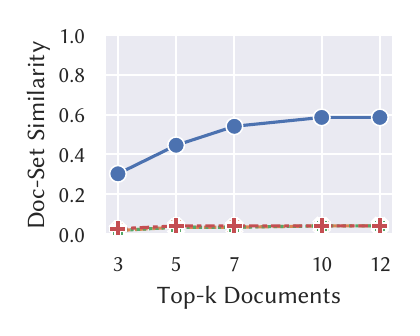}
        \vspace{-0.7cm}
        \caption{\sqldw}
        \label{fig:eval-exact-low-sqldw}
    \end{subfigure}
    \vspace{-0.5cm}
    \caption{
        \captionTitle{Exact Low-Feedback Scenario} Presents the document-set similarity at different top-$k$. The scenario considers the case when a user asks the exact same query as a previously low-rated query. Lower similarity is desired.
    }
        \vspace{-0.5cm}
    \label{fig:eval-exact-low}
\end{figure}

\end{outputfull}

\subsubsection*{Similar High-Feedback}
\label{sec:eval-sim-high}

This scenario considers a user issuing a similar query to one that \system previously answered well. We select queries that have no feedback (i.e., ``unseen'') and are similar to one highly rated query, yielding 55 for \sqldb and 255 for \sqldw. We evaluate using \textbf{Baseline} and \textbf{Feedback} configurations with thresholds $T \in \{0.75, 0.85, 0.95\}$.

\begin{figure}[t!]
    \centering
    \includegraphics[width=\columnwidth]{evalfigures/lowhistory/crop.pdf}
    \begin{subfigure}{0.5\linewidth}
        \includegraphics[width=\linewidth]{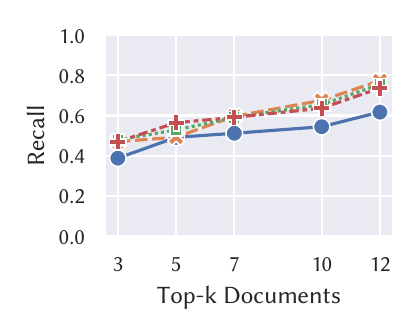}
        \vspace{-0.7cm}
        \caption{\sqldb}
        \label{fig:eval-sim-high-sqldb}
    \end{subfigure}%
    \begin{subfigure}{0.5\linewidth}
        \includegraphics[width=\linewidth]{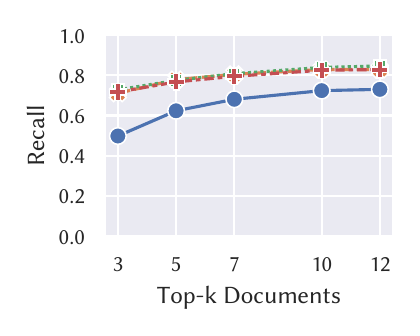}
        \vspace{-0.7cm}
        \caption{\sqldw}
        \label{fig:eval-sim-high-sqldw}
    \end{subfigure}
    \vspace{-0.7cm}
    \caption{
        \captionTitle{Similar High-Feedback} Incorporating feedback signals improves retrieval accuracy.
    }
    \label{fig:eval-sim-high}
    \vspace{-0.6cm}
\end{figure}

Recall is computed against each query's historical reference list to evaluate how well \framework generalizes feedback signals to similar queries. As shown in \cref{fig:eval-sim-high}, \textbf{Feedback(*)} consistently outperforms \textbf{Baseline}. \textbf{Feedback(0.75)} performs best, as its lower threshold admits more chunks into the two-track system (\cref{sec:arch-online}).
Perfect recall is not achieved, as our metric does not fully capture broad-answer scenarios where information may be located across many documents. Nonetheless, \framework enhances \system's ability to retrieve relevant documents by leveraging historical feedback.

\subsubsection*{Unseen Queries}
This scenario considers a user submitting a query that is not similar to previous ones (i.e., ``unseen''). To approximate this, we extract unique queries from historical data, resulting in 1,588 for \sqldb and 3,848 for \sqldw. We evaluate \textbf{Baseline} and \textbf{Feedback} configurations at various thresholds.

\begin{figure}[t!]
    \centering
    \includegraphics[width=\linewidth]{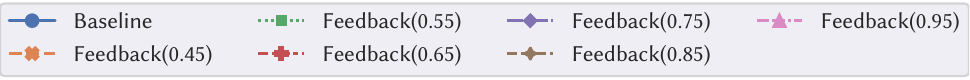}
    \begin{subfigure}{0.5\linewidth}
        \includegraphics[width=\linewidth]{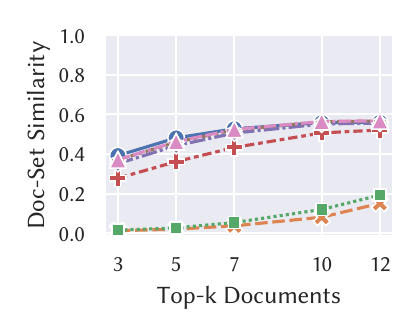}
        \vspace{-0.7cm}
        \caption{\sqldb}
        \label{fig:eval-unknown-history-sqldb}
    \end{subfigure}%
    \begin{subfigure}{0.5\linewidth}
        \includegraphics[width=\linewidth]{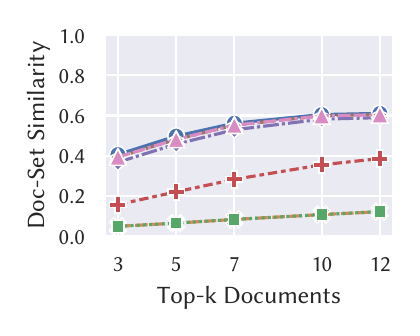}
        \vspace{-0.7cm}
        \caption{\sqldw}
        \label{fig:eval-unknown-history-sqldw}
    \end{subfigure}
    \vspace{-0.7cm}
    \caption{
        \captionTitle{Unseen Queries} \textbf{Feedback} curves that track \textbf{Baseline}'s indicate behavior similar to
        \textbf{Baseline} strategy.
    }
    \label{fig:eval-unknown-history}
    \vspace{-0.3cm}
\end{figure}

We compute similarity against each query’s historical reference list and plot it at various top-$k$ values in \cref{fig:eval-unknown-history}. \textbf{Baseline} reflects the system’s default performance; curves closer to it are preferred, indicating minimal interference.
For both,
higher thresholds (0.75–0.95) produce curves that resemble \textbf{Baseline}'s, while lower thresholds (0.45–0.65) diverge. In the two-track architecture (\cref{sec:arch-online}), the threshold $T$ acts as an ``admit'' filter based on indicator-query similarity (\cref{eq:vote-sig}) that modulates the explore-exploit tradeoff: lowering $T$ admits more low-confidence documents.
Its optimal setting depends on query diversity, knowledge base structure, and tolerance for LLM variability.

Deployed in production, FLAIR also facilitated the evaluation of documentation quality. Over time, reranking patterns revealed outdated documents, controversial contents, and missing details, identifying several documentation gaps across multiple teams.

\begin{outputfull}
\subsection{Scenario: Synthetic Similar High-Rated}

This scenario mimics \cref{sec:eval-sim-high}. However, rather than mining historical
data, we prompt an LLM to generate 5 similar querys from each historical query to
enlarge the test set. This yields 115 querys for \sqldb and 515 querys for
\sqldw. We utilize \chat to generate synthetic querys at different temperature levels (0, 0.05,
0.1, 0.25, 0.5, 1.0), with the goal of obtaining more diverse or creative querys.
We evaluate with the \textbf{Baseline} and \textbf{Feedback} configurations at 0.45, 0.55, 0.65,
0.75, 0.85, and 0.95 thresholds. We assume that each synthetic similar query inherits the same
reference list from the query it was generated from. We plot recall curves at different top-k
values for \sqldb in \cref{fig:eval-synth-sqldb} and for \sqldw in \cref{fig:eval-synth-sqldw}.

\begin{figure}
    \centering
    \includegraphics[width=\linewidth, trim={0 4cm 0 4cm}]{evalfigures/sqldb_synth/sqldb_synth_sim5_legend.pdf}
    \begin{subfigure}{0.33\linewidth}
        \includegraphics[width=\linewidth, trim={0.4cm 0.2cm 0.1cm 0cm}]{evalfigures/sqldb_synth/sqldb_synth_sim5_00.pdf}
        \caption{Temperature = 0.0}
    \end{subfigure}%
    \begin{subfigure}{0.33\linewidth}
        \includegraphics[width=\linewidth, trim={0.4cm 0.2cm 0.1cm 0cm}]{evalfigures/sqldb_synth/sqldb_synth_sim5_05.pdf}
        \caption{Temperature = 0.05}
    \end{subfigure}%
    \begin{subfigure}{0.33\linewidth}
        \includegraphics[width=\linewidth, trim={0.4cm 0.2cm 0.1cm 0cm}]{evalfigures/sqldb_synth/sqldb_synth_sim5_10.pdf}
        \caption{Temperature = 0.1}
    \end{subfigure}

    \begin{subfigure}{0.33\linewidth}
        \includegraphics[width=\linewidth]{evalfigures/sqldb_synth/sqldb_synth_sim5_25.pdf}
        \caption{Temperature = 0.25}
    \end{subfigure}%
    \begin{subfigure}{0.33\linewidth}
        \includegraphics[width=\linewidth, trim={0.4cm 0.2cm 0.3cm 0cm}]{evalfigures/sqldb_synth/sqldb_synth_sim5_50.pdf}
        \caption{Temperature = 0.5}
    \end{subfigure}%
    \begin{subfigure}{0.33\linewidth}
        \includegraphics[width=\linewidth, trim={0.4cm 0.2cm 0.3cm 0cm}]{evalfigures/sqldb_synth/sqldb_synth_sim5_100.pdf}
        \caption{Temperature = 1.0}
    \end{subfigure}
    \caption{
        \captionTitle{\sqldb Similar Synthetic querys} Presents the recall at different top-$k$
        values and thresholds. In this setup, an LLM generates 5
        similar querys from historical highly rated (5-star) querys. We vary the LLM's
        temperature setting to get more varied querys.
    }
    \vspace{2cm}
    \label{fig:eval-synth-sqldb}
\end{figure}

\begin{figure}
    \centering
    \includegraphics[width=\linewidth, trim={0 4cm 0 4cm}]{evalfigures/sqldb_synth/sqldb_synth_sim5_legend.pdf}
    \begin{subfigure}{0.33\linewidth}
        \includegraphics[width=\linewidth, trim={0.4cm 0.2cm 0.1cm 0cm}]{evalfigures/sqldw_synth/sqldb_synth_sim5_00.pdf}
        \caption{Temperature = 0.0}
    \end{subfigure}%
    \begin{subfigure}{0.33\linewidth}
        \includegraphics[width=\linewidth, trim={0.4cm 0.2cm 0.1cm 0cm}]{evalfigures/sqldw_synth/sqldb_synth_sim5_05.pdf}
        \caption{Temperature = 0.05}
    \end{subfigure}%
    \begin{subfigure}{0.33\linewidth}
        \includegraphics[width=\linewidth, trim={0.4cm 0.2cm 0.1cm 0cm}]{evalfigures/sqldw_synth/sqldb_synth_sim5_10.pdf}
        \caption{Temperature = 0.1}
    \end{subfigure}

    \begin{subfigure}{0.33\linewidth}
        \includegraphics[width=\linewidth, trim={0.4cm 0.2cm 0.1cm 0cm}]{evalfigures/sqldw_synth/sqldb_synth_sim5_25.pdf}
        \caption{Temperature = 0.25}
    \end{subfigure}%
    \begin{subfigure}{0.33\linewidth}
        \includegraphics[width=\linewidth, trim={0.4cm 0.2cm 0.1cm 0cm}]{evalfigures/sqldw_synth/sqldb_synth_sim5_50.pdf}
        \caption{Temperature = 0.5}
    \end{subfigure}%
    \begin{subfigure}{0.33\linewidth}
        \includegraphics[width=\linewidth, trim={0.4cm 0.2cm 0.1cm 0cm}]{evalfigures/sqldw_synth/sqldb_synth_sim5_100.pdf}
        \caption{Temperature = 1.0}
    \end{subfigure}
    \caption{
        \captionTitle{\sqldw Similar Synthetic querys} Presents the recall at different top-$k$
        values and thresholds. In this setup, a LLM generates 5
        similar querys from historical highly rated (5-star) querys. We vary the LLM's
        temperature setting to get more varied querys.
    }
    \label{fig:eval-synth-sqldw}
\end{figure}

We observe that for both \sqldb and \sqldw, \textbf{Baseline} exhibits roughly consistent
performance across all temperature levels; generally, we do not observe any strong correlation
between increasing temperature levels and a given configuration's performance.

This experiment illustrates the explore-exploit dynamic of \framework's threshold.
\textbf{Feedback(0.95)} is the most conservative and exhibits the worst performance amongst the
\textbf{Feedback} curves. In certain cases for \sqldw (\cref{fig:eval-synth-sqldw}),
\textbf{Feedback(0.95)} barely improves substantially over \textbf{Baseline}. On the other extreme,
\textbf{Feedback(0.45)} (the most permissive threshold) performs better than
\textbf{Feedback(0.95)} but not substantially for \sqldw. Once again, overly conservative
thresholds restrict retrieval for similar querys that might have slightly larger embedding
distances. On the other end, a very permissive threshold starts pulling in indicators that are
not relevant. We show a sweet spot around \textbf{Feedback(0.65)} on \sqldb
(\cref{fig:eval-synth-sqldb}) and \textbf{Feedback(0.75)} on \sqldw (\cref{fig:eval-synth-sqldw}).

\subsection{Scenario: Reverse HyDE Alignment}
\yz{maybe hide this? or move to appendix?}

We next analyze the impact of generating indicators with synthetic querys on the retrieval of documents relevant to unknown user querys. Specifically, we extract all \textbf{NP} querys (see \cref{sec:eval-data-segment}), representing non-repetitive and unique queries, from historical data. This results in 744 querys for \sqldb and 1,970 querys for \sqldw.

We evaluate these querys using the \textbf{Baseline} and \textbf{RHyDE} configurations from \cref{sec:eval-exact-high}. We define an \textit{impact} as a scenario where \textbf{RHyDE} retrieves a document that \textbf{Baseline} fails to retrieve. If \textbf{RHyDE} retrieves the same document set as \textbf{Baseline}, we consider there to be no impact. For each query where an impact occurs, we manually annotate its \textit{usefulness}, which indicates whether the additional document retrieved by \textbf{RHyDE} contains information relevant to answering the user’s query. This annotation process involves examining the previously retrieved document list and verifying whether the document includes relevant content from the prior answer.

\begin{table}
    \centering
    {\begin{tabular}{c|c|c}
    \toprule
    & \multicolumn{1}{c}{\sqldb} \vline
    & \multicolumn{1}{c}{\sqldw} \\
    \midrule

    Total Unknown & 744 & 1970 \\
    0.75 Threshold Strength/Match & (204) / (N/A) & (430) / (N/A) \\
    0.80 Threshold Strength/Match & (58) / (21) & (89) / (37) \\
    0.85 Threshold Strength/Match & (13) / (9) & (22) / (13) \\
    0.9 Threshold Strength/Match & (1) / (1) & (2) / (0) \\
    \bottomrule
\end{tabular}
}
    \caption{
        \captionTitle{Reverse HyDE Alignment} Demonstrates the impact and usefulness of RHyDE for various thresholds on unknown user querys from \sqldb and \sqldw.
        \textit{Impact} occurs when RHyDE retrieves a document not retrieved by \textbf{Baseline}.
        \textit{Useful} refers to instances where the retrieved document addresses the user’s query.\yz{what is strength and match?}
    }
    \label{tab:rhyde}
\end{table}

The results of this analysis are summarized in \cref{tab:rhyde}. We evaluate four indicator thresholds: 0.75, 0.8, 0.85, and 0.9, which determine the admission of documents in the two-track system (\cref{sec:arch-online}). Due to the scale of data, manual annotations were performed only for thresholds 0.8, 0.85, and 0.9.

From \cref{tab:rhyde}, we observe an impact of approximately 27\% for \sqldb and 21\% for \sqldw at the 0.75 threshold. At the 0.8 threshold, 36\% of impacted querys (58 for \sqldb) and 42\% of impacted querys (89 for \sqldw) are deemed \textit{useful}. This demonstrates that synthetic query generation can effectively enhance retrieval for unknown user querys. We leave the development of curated query sets and dynamic determination of the number of querys per chunk as future work.
\end{outputfull}

\section{Conclusion}


This paper introduced \framework, an efficient and effective framework that integrates feedback learning within 
IR systems, specifically tailored to enhance the adaptiveness of technical copilots. With an offline preprocessing process that generates decentralized dynamic indicators, \framework introduces re-ranking signals into the online retrieval process with a two-track ranking system, addressing several key challenges in retrieval-augmented generation systems, including the static nature of embeddings and the inefficiencies of cache-based systems.
Based on data from deployed production systems, our experimental results underscore the effectiveness of \framework in utilizing both synthetic and real user feedback to refine and evolve retrieval processes. Specifically, we observe accuracy improvements for seen questions (22\%) and unseen questions (10\%), demonstrating \framework's robustness and adaptive handling of novel queries. \framework has now been integrated into the production \system system that serves thousands of \companyname engineers.








\newpage
\balance

\section{GenAI Usage Disclosure}

ChatGPT and GitHub Copilot were utilized to assist in generating sections of this work, including text, tables, graphs, code, data, and citations.

\bibliographystyle{ACM-Reference-Format}
\bibliography{sample}


\begin{thebibliography}{51}


\ifx \showCODEN    \undefined \def \showCODEN     #1{\unskip}     \fi
\ifx \showDOI      \undefined \def \showDOI       #1{#1}\fi
\ifx \showISBNx    \undefined \def \showISBNx     #1{\unskip}     \fi
\ifx \showISBNxiii \undefined \def \showISBNxiii  #1{\unskip}     \fi
\ifx \showISSN     \undefined \def \showISSN      #1{\unskip}     \fi
\ifx \showLCCN     \undefined \def \showLCCN      #1{\unskip}     \fi
\ifx \shownote     \undefined \def \shownote      #1{#1}          \fi
\ifx \showarticletitle \undefined \def \showarticletitle #1{#1}   \fi
\ifx \showURL      \undefined \def \showURL       {\relax}        \fi
\providecommand\bibfield[2]{#2}
\providecommand\bibinfo[2]{#2}
\providecommand\natexlab[1]{#1}
\providecommand\showeprint[2][]{arXiv:#2}

\bibitem[\protect\citeauthoryear{Achiam, Adler, Agarwal, Ahmad, Akkaya, Aleman,
  Almeida, Altenschmidt, Altman, Anadkat, et~al\mbox{.}}{Achiam
  et~al\mbox{.}}{2023}]%
        {gpt}
\bibfield{author}{\bibinfo{person}{Josh Achiam}, \bibinfo{person}{Steven
  Adler}, \bibinfo{person}{Sandhini Agarwal}, \bibinfo{person}{Lama Ahmad},
  \bibinfo{person}{Ilge Akkaya}, \bibinfo{person}{Florencia~Leoni Aleman},
  \bibinfo{person}{Diogo Almeida}, \bibinfo{person}{Janko Altenschmidt},
  \bibinfo{person}{Sam Altman}, \bibinfo{person}{Shyamal Anadkat},
  {et~al\mbox{.}}} \bibinfo{year}{2023}\natexlab{}.
\newblock \showarticletitle{Gpt-4 technical report}.
\newblock \bibinfo{journal}{\emph{arXiv preprint arXiv:2303.08774}}
  (\bibinfo{year}{2023}).
\newblock


\bibitem[\protect\citeauthoryear{Ai, Bi, Guo, and Croft}{Ai
  et~al\mbox{.}}{2018}]%
        {ai2018}
\bibfield{author}{\bibinfo{person}{Qingyao Ai}, \bibinfo{person}{Keping Bi},
  \bibinfo{person}{Jiafeng Guo}, {and} \bibinfo{person}{W.~Bruce Croft}.}
  \bibinfo{year}{2018}\natexlab{}.
\newblock \showarticletitle{Learning a Deep Listwise Context Model for Ranking
  Refinement}. In \bibinfo{booktitle}{\emph{The 41st International ACM SIGIR
  Conference on Research \& Development in Information Retrieval}} (Ann Arbor,
  MI, USA) \emph{(\bibinfo{series}{SIGIR '18})}.
  \bibinfo{publisher}{Association for Computing Machinery},
  \bibinfo{address}{New York, NY, USA}, \bibinfo{pages}{135–144}.
\newblock
\showISBNx{9781450356572}
\urldef\tempurl%
\url{https://doi.org/10.1145/3209978.3209985}
\showDOI{\tempurl}


\bibitem[\protect\citeauthoryear{Anthropic}{Anthropic}{2024}]%
        {claude}
\bibfield{author}{\bibinfo{person}{Anthropic}.}
  \bibinfo{year}{2024}\natexlab{}.
\newblock \bibinfo{booktitle}{\emph{Anthropic Claude}}.
\newblock
\urldef\tempurl%
\url{https://claude.ai/}
\showURL{%
Retrieved Dec 5, 2024 from \tempurl}


\bibitem[\protect\citeauthoryear{Asai, Wu, Wang, Sil, and Hajishirzi}{Asai
  et~al\mbox{.}}{2024}]%
        {asai2024selfrag}
\bibfield{author}{\bibinfo{person}{Akari Asai}, \bibinfo{person}{Zeqiu Wu},
  \bibinfo{person}{Yizhong Wang}, \bibinfo{person}{Avirup Sil}, {and}
  \bibinfo{person}{Hannaneh Hajishirzi}.} \bibinfo{year}{2024}\natexlab{}.
\newblock \showarticletitle{Self-{RAG}: Learning to Retrieve, Generate, and
  Critique through Self-Reflection}. In \bibinfo{booktitle}{\emph{The Twelfth
  International Conference on Learning Representations}}.
\newblock
\urldef\tempurl%
\url{https://openreview.net/forum?id=hSyW5go0v8}
\showURL{%
\tempurl}


\bibitem[\protect\citeauthoryear{Azizi, Meshi, Zoghi, and Karimzadehgan}{Azizi
  et~al\mbox{.}}{2023}]%
        {azizi2023overcomingpriormisspecificationonline}
\bibfield{author}{\bibinfo{person}{Javad Azizi}, \bibinfo{person}{Ofer Meshi},
  \bibinfo{person}{Masrour Zoghi}, {and} \bibinfo{person}{Maryam
  Karimzadehgan}.} \bibinfo{year}{2023}\natexlab{}.
\newblock \bibinfo{title}{Overcoming Prior Misspecification in Online Learning
  to Rank}.
\newblock
\newblock
\showeprint[arxiv]{2301.10651}~[cs.LG]
\urldef\tempurl%
\url{https://arxiv.org/abs/2301.10651}
\showURL{%
\tempurl}


\bibitem[\protect\citeauthoryear{BANERJEE}{BANERJEE}{2005}]%
        {METEOR}
\bibfield{author}{\bibinfo{person}{S. BANERJEE}.}
  \bibinfo{year}{2005}\natexlab{}.
\newblock \showarticletitle{METEOR : An Automatic Metric for MT Evaluation with
  Improved Correlation with Human Judgments}.
\newblock \bibinfo{journal}{\emph{Proceedings of Workshop on Intrinsic and
  Extrinsic Evaluation Measures for MT and/or Summarization at the 43th Annual
  Meeting of the Association of Computational Linguistics (ACL-2005)}}
  (\bibinfo{year}{2005}).
\newblock
\urldef\tempurl%
\url{https://cir.nii.ac.jp/crid/1571698601178637952}
\showURL{%
\tempurl}


\bibitem[\protect\citeauthoryear{Bang}{Bang}{2023}]%
        {bang2023gptcache}
\bibfield{author}{\bibinfo{person}{Fu Bang}.} \bibinfo{year}{2023}\natexlab{}.
\newblock \showarticletitle{GPTCache: An open-source semantic cache for LLM
  applications enabling faster answers and cost savings}. In
  \bibinfo{booktitle}{\emph{Proceedings of the 3rd Workshop for Natural
  Language Processing Open Source Software (NLP-OSS 2023)}}.
  \bibinfo{pages}{212--218}.
\newblock


\bibitem[\protect\citeauthoryear{Bishop and Nasrabadi}{Bishop and
  Nasrabadi}{2006}]%
        {bishop2006pattern}
\bibfield{author}{\bibinfo{person}{Christopher~M Bishop} {and}
  \bibinfo{person}{Nasser~M Nasrabadi}.} \bibinfo{year}{2006}\natexlab{}.
\newblock \bibinfo{booktitle}{\emph{Pattern recognition and machine learning}}.
  Vol.~\bibinfo{volume}{4}.
\newblock \bibinfo{publisher}{Springer}.
\newblock


\bibitem[\protect\citeauthoryear{Burges}{Burges}{1998}]%
        {burges1998tutorial}
\bibfield{author}{\bibinfo{person}{Christopher~JC Burges}.}
  \bibinfo{year}{1998}\natexlab{}.
\newblock \showarticletitle{A tutorial on support vector machines for pattern
  recognition}.
\newblock \bibinfo{journal}{\emph{Data mining and knowledge discovery}}
  \bibinfo{volume}{2}, \bibinfo{number}{2} (\bibinfo{year}{1998}),
  \bibinfo{pages}{121--167}.
\newblock


\bibitem[\protect\citeauthoryear{Cahoon, Singh, Litombe, Larson, Trinh, Zhu,
  Mueller, Psallidas, and Curino}{Cahoon et~al\mbox{.}}{2025}]%
        {joycepaper}
\bibfield{author}{\bibinfo{person}{Joyce Cahoon}, \bibinfo{person}{Prerna
  Singh}, \bibinfo{person}{Nick Litombe}, \bibinfo{person}{Jonathan Larson},
  \bibinfo{person}{Ha Trinh}, \bibinfo{person}{Yiwen Zhu},
  \bibinfo{person}{Andreas Mueller}, \bibinfo{person}{Fotis Psallidas}, {and}
  \bibinfo{person}{Carlo Curino}.} \bibinfo{year}{2025}\natexlab{}.
\newblock \bibinfo{title}{Optimizing open-domain question answering with
  graph-based retrieval augmented generation}.
\newblock
\newblock
\showeprint[arxiv]{2503.02922}~[cs.IR]
\urldef\tempurl%
\url{https://arxiv.org/abs/2503.02922}
\showURL{%
\tempurl}


\bibitem[\protect\citeauthoryear{Campello, Moulavi, and Sander}{Campello
  et~al\mbox{.}}{2013}]%
        {hdbscan}
\bibfield{author}{\bibinfo{person}{Ricardo J. G.~B. Campello},
  \bibinfo{person}{Davoud Moulavi}, {and} \bibinfo{person}{J{\"o}rg Sander}.}
  \bibinfo{year}{2013}\natexlab{}.
\newblock \showarticletitle{Density-Based Clustering Based on Hierarchical
  Density Estimates}. In \bibinfo{booktitle}{\emph{Pacific-Asia Conference on
  Knowledge Discovery and Data Mining}}.
\newblock
\urldef\tempurl%
\url{https://api.semanticscholar.org/CorpusID:32384865}
\showURL{%
\tempurl}


\bibitem[\protect\citeauthoryear{Chang and Zhang}{Chang and Zhang}{2024}]%
        {chang2024communitykgragleveragingcommunitystructures}
\bibfield{author}{\bibinfo{person}{Rong-Ching Chang} {and}
  \bibinfo{person}{Jiawei Zhang}.} \bibinfo{year}{2024}\natexlab{}.
\newblock \bibinfo{title}{CommunityKG-RAG: Leveraging Community Structures in
  Knowledge Graphs for Advanced Retrieval-Augmented Generation in
  Fact-Checking}.
\newblock
\newblock
\showeprint[arxiv]{2408.08535}~[cs.CL]
\urldef\tempurl%
\url{https://arxiv.org/abs/2408.08535}
\showURL{%
\tempurl}


\bibitem[\protect\citeauthoryear{Chen, Lin, Han, and Sun}{Chen
  et~al\mbox{.}}{2023}]%
        {chen2023benchmarkinglargelanguagemodels}
\bibfield{author}{\bibinfo{person}{Jiawei Chen}, \bibinfo{person}{Hongyu Lin},
  \bibinfo{person}{Xianpei Han}, {and} \bibinfo{person}{Le Sun}.}
  \bibinfo{year}{2023}\natexlab{}.
\newblock \bibinfo{title}{Benchmarking Large Language Models in
  Retrieval-Augmented Generation}.
\newblock
\newblock
\showeprint[arxiv]{2309.01431}~[cs.CL]
\urldef\tempurl%
\url{https://arxiv.org/abs/2309.01431}
\showURL{%
\tempurl}


\bibitem[\protect\citeauthoryear{Cormack, Clarke, and Buettcher}{Cormack
  et~al\mbox{.}}{2009}]%
        {rrfpaper}
\bibfield{author}{\bibinfo{person}{Gordon~V. Cormack}, \bibinfo{person}{Charles
  L~A Clarke}, {and} \bibinfo{person}{Stefan Buettcher}.}
  \bibinfo{year}{2009}\natexlab{}.
\newblock \showarticletitle{Reciprocal rank fusion outperforms condorcet and
  individual rank learning methods}. In \bibinfo{booktitle}{\emph{Proceedings
  of the 32nd International ACM SIGIR Conference on Research and Development in
  Information Retrieval}} (Boston, MA, USA) \emph{(\bibinfo{series}{SIGIR
  '09})}. \bibinfo{publisher}{Association for Computing Machinery},
  \bibinfo{address}{New York, NY, USA}, \bibinfo{pages}{758–759}.
\newblock
\showISBNx{9781605584836}
\urldef\tempurl%
\url{https://doi.org/10.1145/1571941.1572114}
\showDOI{\tempurl}


\bibitem[\protect\citeauthoryear{Cover and Hart}{Cover and Hart}{1967}]%
        {cover1967nearest}
\bibfield{author}{\bibinfo{person}{Thomas Cover} {and} \bibinfo{person}{Peter
  Hart}.} \bibinfo{year}{1967}\natexlab{}.
\newblock \showarticletitle{Nearest neighbor pattern classification}.
\newblock \bibinfo{journal}{\emph{IEEE transactions on information theory}}
  \bibinfo{volume}{13}, \bibinfo{number}{1} (\bibinfo{year}{1967}),
  \bibinfo{pages}{21--27}.
\newblock


\bibitem[\protect\citeauthoryear{Dai, Zhao, Ma, Luan, Ni, Lu, Bakalov, Guu,
  Hall, and Chang}{Dai et~al\mbox{.}}{2022}]%
        {dai2022promptagatorfewshotdenseretrieval}
\bibfield{author}{\bibinfo{person}{Zhuyun Dai}, \bibinfo{person}{Vincent~Y.
  Zhao}, \bibinfo{person}{Ji Ma}, \bibinfo{person}{Yi Luan},
  \bibinfo{person}{Jianmo Ni}, \bibinfo{person}{Jing Lu},
  \bibinfo{person}{Anton Bakalov}, \bibinfo{person}{Kelvin Guu},
  \bibinfo{person}{Keith~B. Hall}, {and} \bibinfo{person}{Ming-Wei Chang}.}
  \bibinfo{year}{2022}\natexlab{}.
\newblock \bibinfo{title}{Promptagator: Few-shot Dense Retrieval From 8
  Examples}.
\newblock
\newblock
\showeprint[arxiv]{2209.11755}~[cs.CL]
\urldef\tempurl%
\url{https://arxiv.org/abs/2209.11755}
\showURL{%
\tempurl}


\bibitem[\protect\citeauthoryear{DeepMind}{DeepMind}{2024}]%
        {gemini}
\bibfield{author}{\bibinfo{person}{Google DeepMind}.}
  \bibinfo{year}{2024}\natexlab{}.
\newblock \bibinfo{booktitle}{\emph{Gemini Models}}.
\newblock
\urldef\tempurl%
\url{https://deepmind.google/technologies/gemini/}
\showURL{%
Retrieved November 15, 2024 from \tempurl}


\bibitem[\protect\citeauthoryear{Devlin, Chang, Lee, and Toutanova}{Devlin
  et~al\mbox{.}}{2019}]%
        {bert}
\bibfield{author}{\bibinfo{person}{Jacob Devlin}, \bibinfo{person}{Ming{-}Wei
  Chang}, \bibinfo{person}{Kenton Lee}, {and} \bibinfo{person}{Kristina
  Toutanova}.} \bibinfo{year}{2019}\natexlab{}.
\newblock \showarticletitle{{BERT:} Pre-training of Deep Bidirectional
  Transformers for Language Understanding}. In
  \bibinfo{booktitle}{\emph{Proceedings of the 2019 Conference of the North
  American Chapter of the Association for Computational Linguistics: Human
  Language Technologies, {NAACL-HLT} 2019, Minneapolis, MN, USA, June 2-7,
  2019, Volume 1 (Long and Short Papers)}},
  \bibfield{editor}{\bibinfo{person}{Jill Burstein}, \bibinfo{person}{Christy
  Doran}, {and} \bibinfo{person}{Thamar Solorio}} (Eds.).
  \bibinfo{publisher}{Association for Computational Linguistics},
  \bibinfo{pages}{4171--4186}.
\newblock
\urldef\tempurl%
\url{https://doi.org/10.18653/V1/N19-1423}
\showDOI{\tempurl}


\bibitem[\protect\citeauthoryear{Feng, Feng, Zhao, Yang, and Qin}{Feng
  et~al\mbox{.}}{2024}]%
        {10448015}
\bibfield{author}{\bibinfo{person}{Zhangyin Feng}, \bibinfo{person}{Xiaocheng
  Feng}, \bibinfo{person}{Dezhi Zhao}, \bibinfo{person}{Maojin Yang}, {and}
  \bibinfo{person}{Bing Qin}.} \bibinfo{year}{2024}\natexlab{}.
\newblock \showarticletitle{Retrieval-Generation Synergy Augmented Large
  Language Models}. In \bibinfo{booktitle}{\emph{ICASSP 2024 - 2024 IEEE
  International Conference on Acoustics, Speech and Signal Processing
  (ICASSP)}}. \bibinfo{pages}{11661--11665}.
\newblock
\urldef\tempurl%
\url{https://doi.org/10.1109/ICASSP48485.2024.10448015}
\showDOI{\tempurl}


\bibitem[\protect\citeauthoryear{Gao, Xiong, Gao, Jia, Pan, Bi, Dai, Sun, Wang,
  and Wang}{Gao et~al\mbox{.}}{2024}]%
        {ragsurvey}
\bibfield{author}{\bibinfo{person}{Yunfan Gao}, \bibinfo{person}{Yun Xiong},
  \bibinfo{person}{Xinyu Gao}, \bibinfo{person}{Kangxiang Jia},
  \bibinfo{person}{Jinliu Pan}, \bibinfo{person}{Yuxi Bi}, \bibinfo{person}{Yi
  Dai}, \bibinfo{person}{Jiawei Sun}, \bibinfo{person}{Meng Wang}, {and}
  \bibinfo{person}{Haofen Wang}.} \bibinfo{year}{2024}\natexlab{}.
\newblock \bibinfo{title}{Retrieval-Augmented Generation for Large Language
  Models: A Survey}.
\newblock
\newblock
\showeprint[arxiv]{2312.10997}~[cs.CL]
\urldef\tempurl%
\url{https://arxiv.org/abs/2312.10997}
\showURL{%
\tempurl}


\bibitem[\protect\citeauthoryear{Gon{\c{c}}alves, de~Souza, and
  Gonz{\'a}lez}{Gon{\c{c}}alves et~al\mbox{.}}{2011}]%
        {gonccalves2011collaboration}
\bibfield{author}{\bibinfo{person}{M{\'a}rcio~Kuroki Gon{\c{c}}alves},
  \bibinfo{person}{Cleidson~RB de Souza}, {and} \bibinfo{person}{V{\'\i}ctor~M
  Gonz{\'a}lez}.} \bibinfo{year}{2011}\natexlab{}.
\newblock \showarticletitle{Collaboration, Information Seeking and
  Communication: An Observational Study of Software Developers' Work
  Practices.}
\newblock \bibinfo{journal}{\emph{J. Univers. Comput. Sci.}}
  \bibinfo{volume}{17}, \bibinfo{number}{14} (\bibinfo{year}{2011}),
  \bibinfo{pages}{1913--1930}.
\newblock


\bibitem[\protect\citeauthoryear{Gupta, Ranjan, and Singh}{Gupta
  et~al\mbox{.}}{2024}]%
        {gupta2024comprehensivesurveyretrievalaugmentedgeneration}
\bibfield{author}{\bibinfo{person}{Shailja Gupta}, \bibinfo{person}{Rajesh
  Ranjan}, {and} \bibinfo{person}{Surya~Narayan Singh}.}
  \bibinfo{year}{2024}\natexlab{}.
\newblock \bibinfo{title}{A Comprehensive Survey of Retrieval-Augmented
  Generation (RAG): Evolution, Current Landscape and Future Directions}.
\newblock
\newblock
\showeprint[arxiv]{2410.12837}~[cs.CL]
\urldef\tempurl%
\url{https://arxiv.org/abs/2410.12837}
\showURL{%
\tempurl}


\bibitem[\protect\citeauthoryear{Inc}{Inc}{2024}]%
        {einstein}
\bibfield{author}{\bibinfo{person}{Salesforce Inc}.}
  \bibinfo{year}{2024}\natexlab{}.
\newblock \bibinfo{booktitle}{\emph{How Einstein Copilot Search Uses Retrieval
  Augmented Generation to Make AI More Trusted and Relevant}}.
\newblock
\urldef\tempurl%
\url{https://learn.microsoft.com/en-us/entra/identity/managed-identities-azure-resources/overview}
\showURL{%
Retrieved Dec 5, 2024 from \tempurl}


\bibitem[\protect\citeauthoryear{Jiang, Xu, Gao, Sun, Liu, Dwivedi-Yu, Yang,
  Callan, and Neubig}{Jiang et~al\mbox{.}}{2023}]%
        {jiang-etal-2023-active}
\bibfield{author}{\bibinfo{person}{Zhengbao Jiang}, \bibinfo{person}{Frank Xu},
  \bibinfo{person}{Luyu Gao}, \bibinfo{person}{Zhiqing Sun},
  \bibinfo{person}{Qian Liu}, \bibinfo{person}{Jane Dwivedi-Yu},
  \bibinfo{person}{Yiming Yang}, \bibinfo{person}{Jamie Callan}, {and}
  \bibinfo{person}{Graham Neubig}.} \bibinfo{year}{2023}\natexlab{}.
\newblock \showarticletitle{Active Retrieval Augmented Generation}. In
  \bibinfo{booktitle}{\emph{Proceedings of the 2023 Conference on Empirical
  Methods in Natural Language Processing}},
  \bibfield{editor}{\bibinfo{person}{Houda Bouamor}, \bibinfo{person}{Juan
  Pino}, {and} \bibinfo{person}{Kalika Bali}} (Eds.).
  \bibinfo{publisher}{Association for Computational Linguistics},
  \bibinfo{address}{Singapore}, \bibinfo{pages}{7969--7992}.
\newblock
\urldef\tempurl%
\url{https://doi.org/10.18653/v1/2023.emnlp-main.495}
\showDOI{\tempurl}


\bibitem[\protect\citeauthoryear{Jin, Zhang, Jiang, Liu, Liu, Liu, and Jin}{Jin
  et~al\mbox{.}}{2024}]%
        {jin2024ragcache}
\bibfield{author}{\bibinfo{person}{Chao Jin}, \bibinfo{person}{Zili Zhang},
  \bibinfo{person}{Xuanlin Jiang}, \bibinfo{person}{Fangyue Liu},
  \bibinfo{person}{Xin Liu}, \bibinfo{person}{Xuanzhe Liu}, {and}
  \bibinfo{person}{Xin Jin}.} \bibinfo{year}{2024}\natexlab{}.
\newblock \showarticletitle{RAGCache: Efficient Knowledge Caching for
  Retrieval-Augmented Generation}.
\newblock \bibinfo{journal}{\emph{arXiv preprint arXiv:2404.12457}}
  (\bibinfo{year}{2024}).
\newblock


\bibitem[\protect\citeauthoryear{Kaelbling, Littman, and Moore}{Kaelbling
  et~al\mbox{.}}{1996}]%
        {kaelbling1996reinforcement}
\bibfield{author}{\bibinfo{person}{Leslie~Pack Kaelbling},
  \bibinfo{person}{Michael~L Littman}, {and} \bibinfo{person}{Andrew~W Moore}.}
  \bibinfo{year}{1996}\natexlab{}.
\newblock \showarticletitle{Reinforcement learning: A survey}.
\newblock \bibinfo{journal}{\emph{Journal of artificial intelligence research}}
   \bibinfo{volume}{4} (\bibinfo{year}{1996}), \bibinfo{pages}{237--285}.
\newblock


\bibitem[\protect\citeauthoryear{Karpukhin, Oguz, Min, Lewis, Wu, Edunov, Chen,
  and Yih}{Karpukhin et~al\mbox{.}}{2020}]%
        {karpukhin-etal-2020-dense}
\bibfield{author}{\bibinfo{person}{Vladimir Karpukhin}, \bibinfo{person}{Barlas
  Oguz}, \bibinfo{person}{Sewon Min}, \bibinfo{person}{Patrick Lewis},
  \bibinfo{person}{Ledell Wu}, \bibinfo{person}{Sergey Edunov},
  \bibinfo{person}{Danqi Chen}, {and} \bibinfo{person}{Wen-tau Yih}.}
  \bibinfo{year}{2020}\natexlab{}.
\newblock \showarticletitle{Dense Passage Retrieval for Open-Domain Question
  Answering}. In \bibinfo{booktitle}{\emph{Proceedings of the 2020 Conference
  on Empirical Methods in Natural Language Processing (EMNLP)}},
  \bibfield{editor}{\bibinfo{person}{Bonnie Webber}, \bibinfo{person}{Trevor
  Cohn}, \bibinfo{person}{Yulan He}, {and} \bibinfo{person}{Yang Liu}} (Eds.).
  \bibinfo{publisher}{Association for Computational Linguistics},
  \bibinfo{address}{Online}, \bibinfo{pages}{6769--6781}.
\newblock
\urldef\tempurl%
\url{https://doi.org/10.18653/v1/2020.emnlp-main.550}
\showDOI{\tempurl}


\bibitem[\protect\citeauthoryear{Lewis, Perez, Piktus, Petroni, Karpukhin,
  Goyal, K\"{u}ttler, Lewis, Yih, Rockt\"{a}schel, Riedel, and Kiela}{Lewis
  et~al\mbox{.}}{2020}]%
        {knn}
\bibfield{author}{\bibinfo{person}{Patrick Lewis}, \bibinfo{person}{Ethan
  Perez}, \bibinfo{person}{Aleksandra Piktus}, \bibinfo{person}{Fabio Petroni},
  \bibinfo{person}{Vladimir Karpukhin}, \bibinfo{person}{Naman Goyal},
  \bibinfo{person}{Heinrich K\"{u}ttler}, \bibinfo{person}{Mike Lewis},
  \bibinfo{person}{Wen-tau Yih}, \bibinfo{person}{Tim Rockt\"{a}schel},
  \bibinfo{person}{Sebastian Riedel}, {and} \bibinfo{person}{Douwe Kiela}.}
  \bibinfo{year}{2020}\natexlab{}.
\newblock \showarticletitle{Retrieval-Augmented Generation for
  Knowledge-Intensive NLP Tasks}. In \bibinfo{booktitle}{\emph{Advances in
  Neural Information Processing Systems}},
  \bibfield{editor}{\bibinfo{person}{H.~Larochelle},
  \bibinfo{person}{M.~Ranzato}, \bibinfo{person}{R.~Hadsell},
  \bibinfo{person}{M.F. Balcan}, {and} \bibinfo{person}{H.~Lin}} (Eds.),
  Vol.~\bibinfo{volume}{33}. \bibinfo{publisher}{Curran Associates, Inc.},
  \bibinfo{pages}{9459--9474}.
\newblock
\urldef\tempurl%
\url{https://proceedings.neurips.cc/paper_files/paper/2020/file/6b493230205f780e1bc26945df7481e5-Paper.pdf}
\showURL{%
\tempurl}


\bibitem[\protect\citeauthoryear{Li, Yuan, Feng, Pan, Wang, Sun, Wang, and
  Li}{Li et~al\mbox{.}}{2024b}]%
        {li2024escapeskyhighcostearlystopping}
\bibfield{author}{\bibinfo{person}{Yiwei Li}, \bibinfo{person}{Peiwen Yuan},
  \bibinfo{person}{Shaoxiong Feng}, \bibinfo{person}{Boyuan Pan},
  \bibinfo{person}{Xinglin Wang}, \bibinfo{person}{Bin Sun},
  \bibinfo{person}{Heda Wang}, {and} \bibinfo{person}{Kan Li}.}
  \bibinfo{year}{2024}\natexlab{b}.
\newblock \bibinfo{title}{Escape Sky-high Cost: Early-stopping Self-Consistency
  for Multi-step Reasoning}.
\newblock
\newblock
\showeprint[arxiv]{2401.10480}~[cs.CL]
\urldef\tempurl%
\url{https://arxiv.org/abs/2401.10480}
\showURL{%
\tempurl}


\bibitem[\protect\citeauthoryear{Li, Wang, Jiang, Mao, Chen, Du, Zhang, Zhang,
  Zhang, and Liu}{Li et~al\mbox{.}}{2024a}]%
        {li2024dmqrragdiversemultiqueryrewriting}
\bibfield{author}{\bibinfo{person}{Zhicong Li}, \bibinfo{person}{Jiahao Wang},
  \bibinfo{person}{Zhishu Jiang}, \bibinfo{person}{Hangyu Mao},
  \bibinfo{person}{Zhongxia Chen}, \bibinfo{person}{Jiazhen Du},
  \bibinfo{person}{Yuanxing Zhang}, \bibinfo{person}{Fuzheng Zhang},
  \bibinfo{person}{Di Zhang}, {and} \bibinfo{person}{Yong Liu}.}
  \bibinfo{year}{2024}\natexlab{a}.
\newblock \bibinfo{title}{DMQR-RAG: Diverse Multi-Query Rewriting for RAG}.
\newblock
\newblock
\showeprint[arxiv]{2411.13154}~[cs.IR]
\urldef\tempurl%
\url{https://arxiv.org/abs/2411.13154}
\showURL{%
\tempurl}


\bibitem[\protect\citeauthoryear{OpenAI}{OpenAI}{2024a}]%
        {openaiassistant}
\bibfield{author}{\bibinfo{person}{OpenAI}.} \bibinfo{year}{2024}\natexlab{a}.
\newblock \bibinfo{booktitle}{\emph{Assistants API overview}}.
\newblock
\urldef\tempurl%
\url{https://platform.openai.com/docs/assistants/overview}
\showURL{%
Retrieved Dec 5, 2024 from \tempurl}


\bibitem[\protect\citeauthoryear{OpenAI}{OpenAI}{2024b}]%
        {gpt4o}
\bibfield{author}{\bibinfo{person}{OpenAI}.} \bibinfo{year}{2024}\natexlab{b}.
\newblock \bibinfo{booktitle}{\emph{Hello GPT-4o}}.
\newblock
\urldef\tempurl%
\url{https://openai.com/index/hello-gpt-4o/}
\showURL{%
Retrieved December 5, 2024 from \tempurl}


\bibitem[\protect\citeauthoryear{OpenAI}{OpenAI}{2024c}]%
        {embedding}
\bibfield{author}{\bibinfo{person}{OpenAI}.} \bibinfo{year}{2024}\natexlab{c}.
\newblock \bibinfo{booktitle}{\emph{New embedding models and API updates}}.
\newblock
\urldef\tempurl%
\url{https://openai.com/index/new-embedding-models-and-api-updates/}
\showURL{%
Retrieved July 5, 2024 from \tempurl}


\bibitem[\protect\citeauthoryear{Ouyang, Wu, Jiang, Almeida, Wainwright,
  Mishkin, Zhang, Agarwal, Slama, Ray, Schulman, Hilton, Kelton, Miller,
  Simens, Askell, Welinder, Christiano, Leike, and Lowe}{Ouyang
  et~al\mbox{.}}{2024}]%
        {10.5555/3600270.3602281}
\bibfield{author}{\bibinfo{person}{Long Ouyang}, \bibinfo{person}{Jeff Wu},
  \bibinfo{person}{Xu Jiang}, \bibinfo{person}{Diogo Almeida},
  \bibinfo{person}{Carroll~L. Wainwright}, \bibinfo{person}{Pamela Mishkin},
  \bibinfo{person}{Chong Zhang}, \bibinfo{person}{Sandhini Agarwal},
  \bibinfo{person}{Katarina Slama}, \bibinfo{person}{Alex Ray},
  \bibinfo{person}{John Schulman}, \bibinfo{person}{Jacob Hilton},
  \bibinfo{person}{Fraser Kelton}, \bibinfo{person}{Luke Miller},
  \bibinfo{person}{Maddie Simens}, \bibinfo{person}{Amanda Askell},
  \bibinfo{person}{Peter Welinder}, \bibinfo{person}{Paul Christiano},
  \bibinfo{person}{Jan Leike}, {and} \bibinfo{person}{Ryan Lowe}.}
  \bibinfo{year}{2024}\natexlab{}.
\newblock \showarticletitle{Training language models to follow instructions
  with human feedback}. In \bibinfo{booktitle}{\emph{Proceedings of the 36th
  International Conference on Neural Information Processing Systems}} (New
  Orleans, LA, USA) \emph{(\bibinfo{series}{NIPS '22})}.
  \bibinfo{publisher}{Curran Associates Inc.}, \bibinfo{address}{Red Hook, NY,
  USA}, Article \bibinfo{articleno}{2011}, \bibinfo{numpages}{15}~pages.
\newblock
\showISBNx{9781713871088}


\bibitem[\protect\citeauthoryear{Papineni, Roukos, Ward, and Zhu}{Papineni
  et~al\mbox{.}}{2002}]%
        {10.3115/1073083.1073135}
\bibfield{author}{\bibinfo{person}{Kishore Papineni}, \bibinfo{person}{Salim
  Roukos}, \bibinfo{person}{Todd Ward}, {and} \bibinfo{person}{Wei-Jing Zhu}.}
  \bibinfo{year}{2002}\natexlab{}.
\newblock \showarticletitle{BLEU: a method for automatic evaluation of machine
  translation}. In \bibinfo{booktitle}{\emph{Proceedings of the 40th Annual
  Meeting on Association for Computational Linguistics}} (Philadelphia,
  Pennsylvania) \emph{(\bibinfo{series}{ACL '02})}.
  \bibinfo{publisher}{Association for Computational Linguistics},
  \bibinfo{address}{USA}, \bibinfo{pages}{311–318}.
\newblock
\urldef\tempurl%
\url{https://doi.org/10.3115/1073083.1073135}
\showDOI{\tempurl}


\bibitem[\protect\citeauthoryear{Robertson, Walker, Hancock-Beaulieu, Gatford,
  and Payne}{Robertson et~al\mbox{.}}{1996}]%
        {bm25}
\bibfield{author}{\bibinfo{person}{Stephen Robertson}, \bibinfo{person}{S.
  Walker}, \bibinfo{person}{M.~M. Hancock-Beaulieu}, \bibinfo{person}{M.
  Gatford}, {and} \bibinfo{person}{A. Payne}.} \bibinfo{year}{1996}\natexlab{}.
\newblock \showarticletitle{Okapi at TREC-4}. In \bibinfo{booktitle}{\emph{The
  Fourth Text REtrieval Conference (TREC-4)} (\bibinfo{edition}{the fourth text
  retrieval conference (trec–4)} ed.)}. \bibinfo{publisher}{Gaithersburg, MD:
  NIST}, \bibinfo{pages}{73--96}.
\newblock
\urldef\tempurl%
\url{https://www.microsoft.com/en-us/research/publication/okapi-at-trec-4/}
\showURL{%
\tempurl}


\bibitem[\protect\citeauthoryear{Roy, Zhang, Bhave, Bansal, Las-Casas, Fonseca,
  and Rajmohan}{Roy et~al\mbox{.}}{2024}]%
        {roy2024exploring}
\bibfield{author}{\bibinfo{person}{Devjeet Roy}, \bibinfo{person}{Xuchao
  Zhang}, \bibinfo{person}{Rashi Bhave}, \bibinfo{person}{Chetan Bansal},
  \bibinfo{person}{Pedro Las-Casas}, \bibinfo{person}{Rodrigo Fonseca}, {and}
  \bibinfo{person}{Saravan Rajmohan}.} \bibinfo{year}{2024}\natexlab{}.
\newblock \showarticletitle{Exploring llm-based agents for root cause
  analysis}. In \bibinfo{booktitle}{\emph{Companion Proceedings of the 32nd ACM
  International Conference on the Foundations of Software Engineering}}.
  \bibinfo{pages}{208--219}.
\newblock


\bibitem[\protect\citeauthoryear{Sarthi, Abdullah, Tuli, Khanna, Goldie, and
  Manning}{Sarthi et~al\mbox{.}}{2024}]%
        {sarthi2024raptor}
\bibfield{author}{\bibinfo{person}{Parth Sarthi}, \bibinfo{person}{Salman
  Abdullah}, \bibinfo{person}{Aditi Tuli}, \bibinfo{person}{Shubh Khanna},
  \bibinfo{person}{Anna Goldie}, {and} \bibinfo{person}{Christopher~D
  Manning}.} \bibinfo{year}{2024}\natexlab{}.
\newblock \showarticletitle{{RAPTOR}: Recursive Abstractive Processing for
  Tree-Organized Retrieval}. In \bibinfo{booktitle}{\emph{The Twelfth
  International Conference on Learning Representations}}.
\newblock
\urldef\tempurl%
\url{https://openreview.net/forum?id=GN921JHCRw}
\showURL{%
\tempurl}


\bibitem[\protect\citeauthoryear{Sellam, Das, and Parikh}{Sellam
  et~al\mbox{.}}{2020}]%
        {sellam-etal-2020-bleurt}
\bibfield{author}{\bibinfo{person}{Thibault Sellam}, \bibinfo{person}{Dipanjan
  Das}, {and} \bibinfo{person}{Ankur Parikh}.} \bibinfo{year}{2020}\natexlab{}.
\newblock \showarticletitle{{BLEURT}: Learning Robust Metrics for Text
  Generation}. In \bibinfo{booktitle}{\emph{Proceedings of the 58th Annual
  Meeting of the Association for Computational Linguistics}},
  \bibfield{editor}{\bibinfo{person}{Dan Jurafsky}, \bibinfo{person}{Joyce
  Chai}, \bibinfo{person}{Natalie Schluter}, {and} \bibinfo{person}{Joel
  Tetreault}} (Eds.). \bibinfo{publisher}{Association for Computational
  Linguistics}, \bibinfo{address}{Online}, \bibinfo{pages}{7881--7892}.
\newblock
\urldef\tempurl%
\url{https://doi.org/10.18653/v1/2020.acl-main.704}
\showDOI{\tempurl}


\bibitem[\protect\citeauthoryear{Shi, Zhang, Sun, Gao, Ren, Chen, and Ren}{Shi
  et~al\mbox{.}}{2024}]%
        {shi-etal-2024-generate}
\bibfield{author}{\bibinfo{person}{Zhengliang Shi}, \bibinfo{person}{Shuo
  Zhang}, \bibinfo{person}{Weiwei Sun}, \bibinfo{person}{Shen Gao},
  \bibinfo{person}{Pengjie Ren}, \bibinfo{person}{Zhumin Chen}, {and}
  \bibinfo{person}{Zhaochun Ren}.} \bibinfo{year}{2024}\natexlab{}.
\newblock \showarticletitle{Generate-then-Ground in Retrieval-Augmented
  Generation for Multi-hop Question Answering}. In
  \bibinfo{booktitle}{\emph{Proceedings of the 62nd Annual Meeting of the
  Association for Computational Linguistics (Volume 1: Long Papers)}},
  \bibfield{editor}{\bibinfo{person}{Lun-Wei Ku}, \bibinfo{person}{Andre
  Martins}, {and} \bibinfo{person}{Vivek Srikumar}} (Eds.).
  \bibinfo{publisher}{Association for Computational Linguistics},
  \bibinfo{address}{Bangkok, Thailand}, \bibinfo{pages}{7339--7353}.
\newblock
\urldef\tempurl%
\url{https://doi.org/10.18653/v1/2024.acl-long.397}
\showDOI{\tempurl}


\bibitem[\protect\citeauthoryear{Touvron, Lavril, Izacard, Martinet, Lachaux,
  Lacroix, Rozi{\`e}re, Goyal, Hambro, Azhar, et~al\mbox{.}}{Touvron
  et~al\mbox{.}}{2023}]%
        {llama}
\bibfield{author}{\bibinfo{person}{Hugo Touvron}, \bibinfo{person}{Thibaut
  Lavril}, \bibinfo{person}{Gautier Izacard}, \bibinfo{person}{Xavier
  Martinet}, \bibinfo{person}{Marie-Anne Lachaux},
  \bibinfo{person}{Timoth{\'e}e Lacroix}, \bibinfo{person}{Baptiste
  Rozi{\`e}re}, \bibinfo{person}{Naman Goyal}, \bibinfo{person}{Eric Hambro},
  \bibinfo{person}{Faisal Azhar}, {et~al\mbox{.}}}
  \bibinfo{year}{2023}\natexlab{}.
\newblock \showarticletitle{Llama: Open and efficient foundation language
  models}.
\newblock \bibinfo{journal}{\emph{arXiv preprint arXiv:2302.13971}}
  (\bibinfo{year}{2023}).
\newblock


\bibitem[\protect\citeauthoryear{Trivedi, Balasubramanian, Khot, and
  Sabharwal}{Trivedi et~al\mbox{.}}{2023}]%
        {trivedi-etal-2023-interleaving}
\bibfield{author}{\bibinfo{person}{Harsh Trivedi}, \bibinfo{person}{Niranjan
  Balasubramanian}, \bibinfo{person}{Tushar Khot}, {and}
  \bibinfo{person}{Ashish Sabharwal}.} \bibinfo{year}{2023}\natexlab{}.
\newblock \showarticletitle{Interleaving Retrieval with Chain-of-Thought
  Reasoning for Knowledge-Intensive Multi-Step Questions}. In
  \bibinfo{booktitle}{\emph{Proceedings of the 61st Annual Meeting of the
  Association for Computational Linguistics (Volume 1: Long Papers)}},
  \bibfield{editor}{\bibinfo{person}{Anna Rogers}, \bibinfo{person}{Jordan
  Boyd-Graber}, {and} \bibinfo{person}{Naoaki Okazaki}} (Eds.).
  \bibinfo{publisher}{Association for Computational Linguistics},
  \bibinfo{address}{Toronto, Canada}, \bibinfo{pages}{10014--10037}.
\newblock
\urldef\tempurl%
\url{https://doi.org/10.18653/v1/2023.acl-long.557}
\showDOI{\tempurl}


\bibitem[\protect\citeauthoryear{Wang, Wu, Zhang, Guo, and Zheng}{Wang
  et~al\mbox{.}}{2024}]%
        {WANG2024111334}
\bibfield{author}{\bibinfo{person}{Guojian Wang}, \bibinfo{person}{Faguo Wu},
  \bibinfo{person}{Xiao Zhang}, \bibinfo{person}{Ning Guo}, {and}
  \bibinfo{person}{Zhiming Zheng}.} \bibinfo{year}{2024}\natexlab{}.
\newblock \showarticletitle{Adaptive trajectory-constrained exploration
  strategy for deep reinforcement learning}.
\newblock \bibinfo{journal}{\emph{Knowledge-Based Systems}}
  \bibinfo{volume}{285} (\bibinfo{year}{2024}), \bibinfo{pages}{111334}.
\newblock
\showISSN{0950-7051}
\urldef\tempurl%
\url{https://doi.org/10.1016/j.knosys.2023.111334}
\showDOI{\tempurl}


\bibitem[\protect\citeauthoryear{Ye, Li, Zhang, and Chen}{Ye
  et~al\mbox{.}}{2024}]%
        {ye2024r2agincorporatingretrievalinformation}
\bibfield{author}{\bibinfo{person}{Fuda Ye}, \bibinfo{person}{Shuangyin Li},
  \bibinfo{person}{Yongqi Zhang}, {and} \bibinfo{person}{Lei Chen}.}
  \bibinfo{year}{2024}\natexlab{}.
\newblock \bibinfo{title}{R2AG: Incorporating Retrieval Information into
  Retrieval Augmented Generation}.
\newblock
\newblock
\showeprint[arxiv]{2406.13249}~[cs.CL]
\urldef\tempurl%
\url{https://arxiv.org/abs/2406.13249}
\showURL{%
\tempurl}


\bibitem[\protect\citeauthoryear{Yuan, Neubig, and Liu}{Yuan
  et~al\mbox{.}}{2021}]%
        {NEURIPS2021_e4d2b6e6}
\bibfield{author}{\bibinfo{person}{Weizhe Yuan}, \bibinfo{person}{Graham
  Neubig}, {and} \bibinfo{person}{Pengfei Liu}.}
  \bibinfo{year}{2021}\natexlab{}.
\newblock \showarticletitle{BARTScore: Evaluating Generated Text as Text
  Generation}. In \bibinfo{booktitle}{\emph{Advances in Neural Information
  Processing Systems}}, \bibfield{editor}{\bibinfo{person}{M.~Ranzato},
  \bibinfo{person}{A.~Beygelzimer}, \bibinfo{person}{Y.~Dauphin},
  \bibinfo{person}{P.S. Liang}, {and} \bibinfo{person}{J.~Wortman Vaughan}}
  (Eds.), Vol.~\bibinfo{volume}{34}. \bibinfo{publisher}{Curran Associates,
  Inc.}, \bibinfo{pages}{27263--27277}.
\newblock
\urldef\tempurl%
\url{https://proceedings.neurips.cc/paper_files/paper/2021/file/e4d2b6e6fdeca3e60e0f1a62fee3d9dd-Paper.pdf}
\showURL{%
\tempurl}


\bibitem[\protect\citeauthoryear{Zeighami, Wellmer, and Parameswaran}{Zeighami
  et~al\mbox{.}}{2024}]%
        {zeighami2024nudge}
\bibfield{author}{\bibinfo{person}{Sepanta Zeighami}, \bibinfo{person}{Zac
  Wellmer}, {and} \bibinfo{person}{Aditya Parameswaran}.}
  \bibinfo{year}{2024}\natexlab{}.
\newblock \showarticletitle{NUDGE: Lightweight Non-Parametric Fine-Tuning of
  Embeddings for Retrieval}.
\newblock \bibinfo{journal}{\emph{arXiv preprint arXiv:2409.02343}}
  (\bibinfo{year}{2024}).
\newblock


\bibitem[\protect\citeauthoryear{Zhang, Patil, Jain, Shen, Zaharia, Stoica, and
  Gonzalez}{Zhang et~al\mbox{.}}{2024}]%
        {zhang2024raft}
\bibfield{author}{\bibinfo{person}{Tianjun Zhang}, \bibinfo{person}{Shishir~G
  Patil}, \bibinfo{person}{Naman Jain}, \bibinfo{person}{Sheng Shen},
  \bibinfo{person}{Matei Zaharia}, \bibinfo{person}{Ion Stoica}, {and}
  \bibinfo{person}{Joseph~E Gonzalez}.} \bibinfo{year}{2024}\natexlab{}.
\newblock \showarticletitle{Raft: Adapting language model to domain specific
  rag}.
\newblock \bibinfo{journal}{\emph{arXiv preprint arXiv:2403.10131}}
  (\bibinfo{year}{2024}).
\newblock


\bibitem[\protect\citeauthoryear{Zhang, Li, Cui, Cai, Liu, Fu, Huang, Zhao,
  Zhang, Chen, Wang, Luu, Bi, Shi, and Shi}{Zhang et~al\mbox{.}}{2023}]%
        {zhang2023sirenssongaiocean}
\bibfield{author}{\bibinfo{person}{Yue Zhang}, \bibinfo{person}{Yafu Li},
  \bibinfo{person}{Leyang Cui}, \bibinfo{person}{Deng Cai},
  \bibinfo{person}{Lemao Liu}, \bibinfo{person}{Tingchen Fu},
  \bibinfo{person}{Xinting Huang}, \bibinfo{person}{Enbo Zhao},
  \bibinfo{person}{Yu Zhang}, \bibinfo{person}{Yulong Chen},
  \bibinfo{person}{Longyue Wang}, \bibinfo{person}{Anh~Tuan Luu},
  \bibinfo{person}{Wei Bi}, \bibinfo{person}{Freda Shi}, {and}
  \bibinfo{person}{Shuming Shi}.} \bibinfo{year}{2023}\natexlab{}.
\newblock \bibinfo{title}{Siren's Song in the AI Ocean: A Survey on
  Hallucination in Large Language Models}.
\newblock
\newblock
\showeprint[arxiv]{2309.01219}~[cs.CL]
\urldef\tempurl%
\url{https://arxiv.org/abs/2309.01219}
\showURL{%
\tempurl}


\bibitem[\protect\citeauthoryear{Zhou, Zhang, Hasson, Singh, and Li}{Zhou
  et~al\mbox{.}}{2024}]%
        {rhyde}
\bibfield{author}{\bibinfo{person}{Weichao Zhou}, \bibinfo{person}{Jiaxin
  Zhang}, \bibinfo{person}{Hilaf Hasson}, \bibinfo{person}{Anu Singh}, {and}
  \bibinfo{person}{Wenchao Li}.} \bibinfo{year}{2024}\natexlab{}.
\newblock \showarticletitle{{H}y{QE}: Ranking Contexts with Hypothetical Query
  Embeddings}. In \bibinfo{booktitle}{\emph{Findings of the Association for
  Computational Linguistics: EMNLP 2024}},
  \bibfield{editor}{\bibinfo{person}{Yaser Al-Onaizan}, \bibinfo{person}{Mohit
  Bansal}, {and} \bibinfo{person}{Yun-Nung Chen}} (Eds.).
  \bibinfo{publisher}{Association for Computational Linguistics},
  \bibinfo{address}{Miami, Florida, USA}, \bibinfo{pages}{13014--13032}.
\newblock
\urldef\tempurl%
\url{https://doi.org/10.18653/v1/2024.findings-emnlp.761}
\showDOI{\tempurl}


\bibitem[\protect\citeauthoryear{Zhu, Sheng, Zheng, Barrett, Jordan, and
  Jiao}{Zhu et~al\mbox{.}}{2024}]%
        {zhu2024towards}
\bibfield{author}{\bibinfo{person}{Banghua Zhu}, \bibinfo{person}{Ying Sheng},
  \bibinfo{person}{Lianmin Zheng}, \bibinfo{person}{Clark Barrett},
  \bibinfo{person}{Michael Jordan}, {and} \bibinfo{person}{Jiantao Jiao}.}
  \bibinfo{year}{2024}\natexlab{}.
\newblock \showarticletitle{Towards Optimal Caching and Model Selection for
  Large Model Inference}.
\newblock \bibinfo{journal}{\emph{Advances in Neural Information Processing
  Systems}}  \bibinfo{volume}{36} (\bibinfo{year}{2024}).
\newblock


\bibitem[\protect\citeauthoryear{Zhu, Demarne, Deng, Wang, Sahoo, Vermareddy,
  Lerner, Lu, Bararia, Bhavan, Zhang, Li, Lin, Cilimdzic, and Krishnan}{Zhu
  et~al\mbox{.}}{2025}]%
        {dricopilot}
\bibfield{author}{\bibinfo{person}{Yiwen Zhu}, \bibinfo{person}{Mathieu
  Demarne}, \bibinfo{person}{Kai Deng}, \bibinfo{person}{Wenjing Wang},
  \bibinfo{person}{Nutan Sahoo}, \bibinfo{person}{Divya Vermareddy},
  \bibinfo{person}{Hannah Lerner}, \bibinfo{person}{Yunlei Lu},
  \bibinfo{person}{Swati Bararia}, \bibinfo{person}{Anjali Bhavan},
  \bibinfo{person}{William Zhang}, \bibinfo{person}{Xia Li},
  \bibinfo{person}{Katherine Lin}, \bibinfo{person}{Miso Cilimdzic}, {and}
  \bibinfo{person}{Subru Krishnan}.} \bibinfo{year}{2025}\natexlab{}.
\newblock \bibinfo{title}{DECO: Life-Cycle Management of Enterprise-Grade
  Copilots}.
\newblock
\newblock
\showeprint[arxiv]{2412.06099}~[cs.SE]
\urldef\tempurl%
\url{https://arxiv.org/abs/2412.06099}
\showURL{%
\tempurl}


\end{thebibliography}

\end{document}